\documentclass[preprint,12pt]{elsarticle}

 \usepackage{graphicx}

\usepackage{amssymb}

\journal{Nuclear Physics A}

\begin{document}

\begin{frontmatter}



\title{Variational study for the equation of state \\
of asymmetric nuclear matter at finite temperatures}


\author[WU]{H.~Togashi\corref{cor1}}
\ead{hajime\_togashi@ruri.waseda.jp}

\author[WU,RISE]{M.~Takano}

\cortext[cor1]{Corresponding author}
\address[WU]{Department of Pure and Applied Physics, 
Graduate School of Advanced Science and Engineering, 
Waseda University, 3-4-1 Okubo Shinjuku-ku, Tokyo 169-8555, Japan}
\address[RISE]{Research Institute for Science and Engineering, 
Waseda University, 3-4-1 Okubo Shinjuku-ku, Tokyo 169-8555, Japan}

\begin{abstract}
An equation of state (EOS) for uniform asymmetric nuclear matter (ANM) is constructed at zero and finite temperatures by the variational method starting from the nuclear Hamiltonian that is composed of the Argonne v18 and Urbana IX potentials. 
At zero temperature, the two-body energy is calculated with the Jastrow wave function in the two-body cluster approximation which is supplemented by Mayer's condition and the healing-distance condition so as to reproduce the result by Akmal, Pandharipande and Ravenhall. 
The energy caused by the three-body force is treated somewhat phenomenologically so that the total energy reproduces the empirical saturation conditions. 
The masses and radii of neutron stars obtained with the EOS are consistent with recent observational data. 
At finite temperatures, thermodynamic quantities such as free energy, internal energy, entropy, pressure and chemical potentials are calculated with an extension of the method by Schmidt and Pandharipande.  
The validity of the frozen-correlation approximation employed in this work is confirmed as compared with the result of the fully minimized calculation.  
The quadratic proton-fraction-dependence of the energy of ANM is confirmed at zero temperature, whereas the free energy of ANM deviates from the quadratic proton-fraction-dependence markedly at finite temperatures.  
The obtained EOS of ANM will be an important ingredient of a new nuclear EOS for supernova numerical simulations.  

\end{abstract}

\begin{keyword}
Nuclear matter \sep Nuclear EOS \sep Variational method \sep Neutron stars \sep Supernovae

\end{keyword}

\end{frontmatter}


\section{Introduction}
The nuclear equation of state (EOS) plays important roles in the studies of high-energy astrophysical phenomena: 
The nuclear EOS at zero temperature governs the structure of cold neutron stars (NSs), whereas the EOS for finite temperatures is necessary for studies of core collapse supernovae (SNe), black hole formations, cooling of NSs and so on. 
At zero temperature, a variety of nuclear EOSs, including those based on the realistic nuclear Hamiltonian, have been applied to NSs, and recent observational data on NSs impose severe constraints on the EOS \cite{2Msolar,Steiner}.  
On the other hand, the nuclear EOSs available for the numerical simulations of SN are limited, because it is a hard task to construct an EOS for the SN simulations:
The SN-EOS table must provide thermodynamic quantities in an extremely wide range of densities, temperatures and proton fractions.  
Furthermore, the EOS of non-uniform nuclear matter as well as that of uniform matter is important in SN simulations.  

For the above reasons, until several years ago, there were only two kinds of nuclear EOSs available for SN simulations. 
One was constructed by Lattimer and Swesty \cite{LS} with use of a Skyrme-type interaction for uniform matter and a compressible liquid-drop model for non-uniform matter. 
The other was constructed by Shen et al. \cite{Shen} with use of a relativistic mean field (RMF) theory for uniform matter and a Thomas-Fermi (TF) approximation for non-uniform matter.  
Recently, new SN-EOSs based on the RMF theory have been proposed: 
For the high-density region, the Shen-EOS was extended so as to take into account hyperon mixing \cite{Ishizuka} and the quark-hadron phase transition \cite{Nakazato}.  
Furthermore, the RMF calculations with parameters different from those of the Shen-EOS have been applied to SN-EOSs \cite{Hempel, GShen}. 
For the low-density region, the nuclear statistical equilibrium is taken into account \cite{Hempel, Furusawa}, or the virial expansion is adopted \cite{GShen}.  

These recent advances in the SN-EOSs are astonishing.  
However, in these EOSs, uniform matter is treated with phenomenological nuclear models: 
To our knowledge, there are no EOSs based on the nuclear Hamiltonian composed of bare nuclear forces.  
This situation motivates us to construct a new nuclear EOS for SN simulations based on a realistic nuclear Hamiltonian with use of the variational many-body theory.  

In the standard variational method, the Jastrow wave function is assumed, and the expectation value of the Hamiltonian is cluster-expanded.  
Then, parts of the higher-order cluster terms are resummed by the Fermi Hypernetted Chain (FHNC) method \cite{Clark}.  
Akmal, Pandharipande and Ravenhall (APR) performed the FHNC calculation \cite{APR} with the Argonne v18 (AV18) two-body potential \cite{AV18} and the Urbana IX (UIX) three-body potential \cite{UIX}; the obtained EOS has been referred to as one of the standard nuclear EOSs at zero temperature.  
In Ref. \cite{AM}, the EOS of APR is extended to finite temperatures with the method proposed by Schmidt and Pandharipande (SP) \cite{SP, MP}. 
Furthermore, the quantum Monte Carlo (QMC) techniques have been developed for nuclear matter at zero temperature \cite{GFMC, AFDMC}, and, in Ref. \cite{Illarionov}, free energy of hot nuclear matter is calculated with the method of SP in connection with the EOS by one of the QMC methods.  
However, these variational methods are applied only to limited cases: symmetric nuclear matter (SNM) with the proton fraction $x$ being $x=1/2$ and pure neutron matter (PNM) with $x=0$.  
This is because the energy of asymmetric nuclear matter (ANM) with 0$< x <$1/2 is difficult to calculate by those variational methods.  
In fact, there are only few examples of variational calculations for ANM \cite{LP, LOCV1, Takano}.  
Furthermore, sophisticated numerical many-body calculations are, practically, not suitable for constructing an SN-EOS, because, including the non-uniform phase, hundreds of thousands of data on thermodynamic quantities must be completed in an SN-EOS table.  

In our project, we adopt a relatively simple variational method that can treat ANM, and have so far constructed a realistic nuclear EOS for SNM and PNM in Ref. \cite{K1}, hereafter referred to as paper I.  
In that paper, starting from the nuclear Hamiltonian with the AV18 and UIX potentials, we calculated the two-body energy in the two-body-cluster approximation, 
supplemented by appropriate constraints, and incorporated the three-body energy somewhat phenomenologically so that the total energy could reproduce the empirical saturation data. 
At finite temperatures, we employed the method of SP to obtain the free energy. 

In order to complete an SN-EOS, it is necessary to treat non-uniform matter as well as uniform matter.  
Since we focus on improving the treatment of uniform matter using the variational method, we simply adopt the TF calculation for non-uniform matter, following the method by Shen et al \cite{Shen}.  
Here, we take into account the isolated atomic nuclei:   
Since a large amount of experimental data on the atomic nuclei are available, we fine-tuned the EOS so that the TF calculation for the atomic nuclei could reproduce the gross features of their masses and radii in Ref. \cite{K2}, hereafter referred to as paper II.  
In that paper, we further constructed a non-uniform $\beta$-stable EOS at zero temperature, which corresponds to the EOS of the NS crust, by the TF calculation.  
The obtained EOS is consistent with typical EOSs calculated for NS crusts \cite{BPS, NV}.  

Toward an SN-EOS, the next important task is to construct the EOS of uniform ANM with $0 < x < 1/2$, which is the main object of this paper.  
Namely, we extend the variational method in papers I and II to obtain the EOS of uniform ANM at zero and finite temperatures.  
In Section 2, we perform this extension for nuclear matter at zero temperature, and calculate the EOS of ANM.  
Then, we apply the EOS to NSs, and compare the NS solutions with recent observational data.  
In addition, we discuss the $x$-dependence of the energy of ANM quantitatively.  
In Section 3, we treat ANM at finite temperatures to calculate various thermodynamic quantities of ANM.  
We also examine the validity of the so-called frozen-correlation approximation adopted in our study: 
For this purpose, we perform the full minimization of the free energy in the present variational method.  
Furthermore, we discuss the $x$-dependence of the free energy of ANM at finite temperatures.  
A summary is given in Section 4.

\section{Asymmetric nuclear matter at zero temperature}
\subsection{Extension of the variational method to asymmetric nuclear matter}
In this section, we calculate the energy per nucleon $E/N$ of uniform ANM as a function of the total nucleon number density $\rho$ and the proton fraction $x$.  
As in papers I and II, we start from the non-relativistic nuclear Hamiltonian $H$ for $N$ nucleons interacting through realistic two-body and three-body forces, and decompose $H$ into the two-body Hamiltonian $H_2$ and the three-body Hamiltonian $H_3$ as follows: 
\begin{equation}
H = H_2 + H_3.
\end{equation}

The two-body Hamiltonian $H_2$ is written as 
\begin{equation}
H_2 = -{\textstyle\sum\limits_{i=1}^N} \frac{\hbar^2}{2m}\nabla_i^2
      + {\textstyle\sum\limits_{i<j}^N}V_{ij},      \label{eq:H2}
\end{equation}
where $m$ is the nucleon mass: For simplicity, we choose $m$ in Eq. (\ref{eq:H2}) as the mass of a neutron, as in paper II.  
For the two-body nuclear potential $V_{ij}$ acting on an $(i, j)$ nucleon pair, we use the isoscalar part of the AV18 potential as in papers I and II, namely, 
\begin{eqnarray}
V_{ij} &=& {\textstyle\sum\limits_{t=0}^1}{\textstyle\sum\limits_{s=0}^1}
        [V_{\mathrm{C}ts}(r_{ij}) + sV_{\mathrm{T}t}(r_{ij})S_{\mathrm{T}ij} 
+ sV_{\mathrm{SO}t}(r_{ij})(\mbox{\boldmath$L$}_{ij}\cdot\mbox{\boldmath$s$})  \nonumber   \\
       && + V_{\mathrm{qL}ts}(r_{ij})
\left| \mbox{\boldmath$L$}_{ij} \right| ^2
 + sV_{\mathrm{qSO}t}(r_{ij})(\mbox{\boldmath$L$}_{ij}\cdot\mbox{\boldmath$s$})^2]P_{tsij}.
\end{eqnarray}
Here, $S_{\mathrm{T}ij}$ is the tensor operator, 
$\mbox{\boldmath$L$}_{ij}$ is the relative orbital angular momentum operator, 
and $P_{tsij}$ is the spin-isospin projection operator projecting a two-nucleon state onto the spin-isospin eigenstates with $t$ and $s$ being the two-nucleon total isospin and spin, respectively.

The three-body Hamiltonian $H_3$ is given by 
\begin{equation}
H_3={\textstyle\sum\limits_{i<j<k}^N}V_{ijk}. 
\end{equation}
As the three-body nuclear potential $V_{ijk}$, we employ the UIX potential,
which consists of the repulsive component $V_{ijk}^{\mathrm{R}}$ and the two-pion exchange component $V_{ijk}^{2\pi}$:
\begin{equation}
V_{ijk}=V_{ijk}^{\mathrm{R}}+V_{ijk}^{2\pi}.   \label{eq:UIX}
\end{equation}

We first calculate the two-body energy $E_2/N$ without the three-body force.  
As in papers I and II, we use the two-body cluster approximation for the expectation value of $H_2$ with the following Jastrow wave function: 
\begin{equation}
\mathnormal{\Psi}=\mathrm{Sym}\left[\prod_{i<j}f_{ij}\right]\mathnormal{\Phi}_\mathrm{F}.  \label{eq:Jastrow}
\end{equation}
Here, $\mathnormal{\Phi}_\mathrm{F}$ is the degenerate Fermi-gas wave function at zero temperature and $\mathrm{Sym}$ is the symmetrizer with respect to the order of the factors in the products. 
The two-body correlation function $f_{ij}$ is taken as
\begin{equation}
f_{ij}={\textstyle\sum\limits_{t=0}^1}{\textstyle\sum\limits_{\mu}}{\textstyle\sum\limits_{s=0}^1}
        \left[f_{\mathrm{C}ts}^{\mu}(r_{ij})+sf_{\mathrm{T}t}^{\mu}(r_{ij})S_{\mathrm{T}ij}
        +sf_{\mathrm{SO}t}^{\mu}(r_{ij})
        (\mbox{\boldmath$L$}_{ij}\cdot\mbox{\boldmath$s$})\right]P_{tsij}^{\mu}.     \label{eq:fij}
\end{equation}
The Jastrow wave function of ANM in Eq. (\ref{eq:Jastrow}) differs from those of symmetric nuclear matter (SNM) and of pure neutron matter (PNM) 
in the following ways: 
i) The Fermi wave numbers of protons $k_{\mathrm{Fp}}=(3\pi\rho x)^{1/3}$ and neutrons $k_{\mathrm{Fn}}=[3\pi\rho (1-x)]^{1/3}$
 included in $\mathnormal{\Phi}_\mathrm{F}$  are in general different from each other, depending on $x$.  
ii) The correlation function $f_{ij}$ depends on the third component of the isospin $t_3$: 
Even though the two-body interaction $V_{ij}$ is isoscalar, we expect that the two-body correlations for isospin-triplet three states (p-p, p-n, n-n) may differ from each other in, e.g., extremely neutron-rich ANM, due to an effect of the medium.  
In Eq. (\ref{eq:fij}), the $t_3$-dependence of $f_{ij}$ is expressed with the superscript $\mu$, i.e., $\mu = (+, 0, -)$ for $t_3 =(1, 0, -1)$.    
Namely, the central, tensor and spin-orbit correlation functions $f_{\mathrm{C}ts}^{\mu}(r_{ij})$, $f_{\mathrm{T}t}^{\mu}(r_{ij})$ 
and $f_{\mathrm{SO}t}^{\mu}(r_{ij})$ depend not only on $(t, s)$ but also on $\mu$.  
In addition, the projection operator $P_{tsij}^{\mu}$ projects the $(i, j)$ nucleon pair states onto the $\mu=(+, 0, -)$ states.  
Therefore, in the case of ANM, these 16 correlation functions are regarded as variational functions.

With use of this trial wave function $\mathnormal{\Psi}$, we express the expectation value of $H_2$ in the two-body cluster approximation to obtain $E_2/N$: 
\begin{eqnarray}
\frac{E_2(\rho, x)}{N} &=& \frac{E_1}{N} +2\pi\rho{\textstyle\sum\limits_{t=0}^1}{\textstyle\sum\limits_{\mu}}{\textstyle\sum\limits_{s=0}^1}
\int_0^{\infty}r^2dr\bigg[F_{ts}^{\mu}(r)V_{\mathrm{C}ts}(r)+sF_{\mathrm{T}t}^{\mu}(r)V_{\mathrm{T}t}(r)   \nonumber  \\
   && +sF_{\mathrm{SO}t}^{\mu}(r)V_{\mathrm{SO}t}(r)+
F_{\mathrm{qL}ts}^{\mu}(r)V_{\mathrm{qL}ts}(r)+sF_{\mathrm{qSO}t}^{\mu}(r)V_{\mathrm{qSO}t}(r)\bigg]  \nonumber  \\
   && +\frac{2\pi\rho\hbar^2}{m}{\textstyle\sum\limits_{t=0}^1}{\textstyle\sum\limits_{\mu}}{\textstyle\sum\limits_{s=0}^1}\int_0^{\infty}r^2dr
\Bigg[\Bigg\{\left[\frac{df_{\mathrm{C}ts}^{\mu}(r)}{dr}\right]^2+8s\left[\frac{df_{\mathrm{T}t}^{\mu}(r)}{dr}\right]^2  \nonumber  \\
   && +48s\left[\frac{f_{\mathrm{T}t}^{\mu}(r)}{r}\right]^2\Bigg\}F_{\mathrm{F}ts}^{\mu}(r)+\frac{2}{3}s\left[\frac{df_{\mathrm{SO}t}^{\mu}(r)}{dr}\right]^2
F_{\mathrm{qF}ts}^{\mu}(r)\Bigg].   \label{eq:E2}
\end{eqnarray}
On the right-hand side of Eq. (\ref{eq:E2}), the first term $E_1/N$ is the one-body kinetic energy given by
\begin{equation}
\frac{E_1(\rho, x)}{N}=\frac{3}{5}\bigg[x\frac{\hbar^2k_{\mathrm{Fp}}^2}{2m}+(1-x)\frac{\hbar^2k_{\mathrm{Fn}}^2}{2m}\bigg].   \label{eq:FermiE}
\end{equation}
The second term represents the potential energy, where $F_{ts}^{\mu}(r)$, $F_{\mathrm{T}t}^{\mu}(r)$, 
$F_{\mathrm{SO}t}^{\mu}(r)$, $F_{\mathrm{qL}ts}^{\mu}(r)$ and $F_{\mathrm{qSO}t}^{\mu}(r)$ are defined as
\begin{eqnarray}
F_{ts}^{\mu}(r) = \bigg\{\left[f_{\mathrm{C}ts}^{\mu}(r)\right]^2+8s\left[f_{\mathrm{T}t}^{\mu}(r)\right]^2\bigg\}F_{\mathrm{F}ts}^{\mu}(r)
+\frac{2}{3}s\left[f_{\mathrm{SO}t}^{\mu}(r)\right]^2F_{\mathrm{qF}ts}^{\mu}(r),   \label{eq:Fts}
\end{eqnarray}
\begin{equation}
F_{\mathrm{T}t}^{\mu}(r) = 16f_{\mathrm{T}t}^{\mu}(r)\left[f_{\mathrm{C}t1}^{\mu}(r)-f_{\mathrm{T}t}^{\mu}(r)\right]F_{\mathrm{F}t1}^{\mu}(r)
-\frac{2}{3}\left[f_{\mathrm{SO}t}^{\mu}(r)\right]^2F_{\mathrm{qF}t1}^{\mu}(r),    
\end{equation}
\begin{eqnarray}
F_{\mathrm{SO}t}^{\mu}(r) &=& \Bigg\{\frac{4}{3}f_{\mathrm{SO}t}^{\mu}(r)\left[f_{\mathrm{C}t1}^{\mu}(r)-f_{\mathrm{T}t}^{\mu}(r)\right]
-\frac{1}{3}\left[f_{\mathrm{SO}t}^{\mu}(r)\right]^2\Bigg\}F_{\mathrm{qF}t1}^{\mu}(r)   \nonumber \\
&&-24\left[f_{\mathrm{T}t}^{\mu}(r)\right]^2F_{\mathrm{F}t1}^{\mu}(r),     
\end{eqnarray}
\begin{eqnarray}
F_{\mathrm{qL}ts}^{\mu}(r) &=& 48s\left[f_{\mathrm{T}t}^{\mu}(r)\right]^2F_{\mathrm{F}ts}^{\mu}(r)+\bigg\{\left[f_{\mathrm{C}ts}^{\mu}(r)\right]^2
+8s\left[f_{\mathrm{T}t}^{\mu}(r)\right]^2\bigg\}F_{\mathrm{qF}ts}^{\mu}(r)     \nonumber  \\
&&+\frac{2}{3}s\left[f_{\mathrm{SO}t}^{\mu}(r)\right]^2F_{\mathrm{bF}ts}^{\mu}(r),    
\end{eqnarray}
\begin{eqnarray}
F_{\mathrm{qSO}t}^{\mu}(r) &=& 72\left[f_{\mathrm{T}t}^{\mu}(r)\right]^2F_{\mathrm{F}t1}^{\mu}(r)+\Bigg\{\frac{2}{3}\left[f_{\mathrm{C}t1}^{\mu}(r)\right]^2
-\frac{4}{3}f_{\mathrm{C}t1}^{\mu}(r)f_{\mathrm{T}t}^{\mu}(r)     \nonumber   \\
  && -\frac{2}{3}f_{\mathrm{C}t1}^{\mu}(r)f_{\mathrm{SO}t}^{\mu}(r)
+\frac{20}{3}\left[f_{\mathrm{T}t}^{\mu}(r)\right]^2+\frac{8}{3}f_{\mathrm{T}t}^{\mu}(r)f_{\mathrm{SO}t}^{\mu}(r)\Bigg\}F_{\mathrm{qF}t1}^{\mu}(r)  \nonumber \\
&&+\frac{2}{3}\left[f_{\mathrm{SO}t}^{\mu}(r)\right]^2F_{\mathrm{bF}t1}^{\mu}(r).      \label{eq:Fqsot}
\end{eqnarray}
Here, $F_{\mathrm{F}ts}^{\mu}(r)$, $F_{\mathrm{qF}ts}^{\mu}(r)$ and $F_{\mathrm{bF}ts}^{\mu}(r)$ are various distribution functions in the case of the degenerate Fermi gas, the definitions and explicit expressions of which are given in Appendix A. 
We note that these expressions are similar to those for SNM and PNM shown in paper I, though the $x$-dependence is introduced into the relevant functions with the superscript $\mu$.  

Then, we minimize $E_2/N$ with respect to  $f_{\mathrm{C}ts}^{\mu}(r)$, $f_{\mathrm{T}t}^{\mu}(r)$ 
and $f_{\mathrm{SO}t}^{\mu}(r)$ by solving the Euler-Lagrange equations.  
Here, as in papers I and II, we impose two constraints in order to compensate effectively for the neglect of higher-order cluster terms.  

The first constraint is the extended Mayer's condition \cite{K1, Mayer}, which is a kind of normalization condition, expressed as
\begin{equation}
\rho\int_0^{\infty}\left[F_{ts}^{\mu}(r)-F_{\mathrm{F}ts}^{\mu}(r)\right]d\mbox{\boldmath$r$}=0.  \label{eq:mayer}
\end{equation}
Here, $F_{ts}^{\mu}(r)$ is a two-body cluster approximation of the $(t, s, \mu)$-projected radial distribution function shown in Eq. (\ref{eq:Fts}). 
When Eq. (\ref{eq:mayer}) is summed over $\mu$, this condition reduces to that adopted for SNM and PNM in papers I and II.  

The second constraint is the healing-distance condition expressed as
\begin{equation}
f_{\mathrm{C}ts}^{\mu}(r)=1, \ \ \ \ \ \ \ f_{\mathrm{T}t}^{\mu}(r)=0, \ \ \ \ \ \ \ f_{\mathrm{SO}t}^{\mu}(r)=0\ \ \ \ \ \ \ (r \geq r_\mathrm{h}),
\end{equation}
which means that the correlation between two nucleons vanishes at the distances $r$ that are larger than the healing distance $r_{\mathrm{h}}$.
Here, we assume that $r_{\mathrm{h}}$ is common to the central, tensor and spin-orbit correlation functions, and proportional to the mean distance between nucleons, i.e., $r_{\mathrm{h}}=a_{\mathrm{h}}r_0$ with $r_0=(3/4\pi\rho)^{1/3}$. 
The coefficient is chosen to be $a_{\mathrm{h}}=1.76$ so that $E_2/N$ reproduces the result of the FHNC calculation by APR \cite{APR} which includes the higher-order cluster terms.  
As shown in paper I, the obtained $E_2/N$ with $a_{\mathrm{h}}=1.76$ for both SNM and PNM are in good agreement with those by APR, and, in this paper, we assume that $a_{\mathrm{h}}$ is independent of $x$.  

Next, we calculate the three-body energy $E_3/N$, based on the three-body Hamiltonian $H_3$. 
Since there is some uncertainty in the three-body force, we express $E_3/N$ somewhat phenomenologically as 
\begin{equation}
\frac{E_3(\rho, x)}{N}=\frac{1}{N} \langle {\textstyle\sum\limits_{i<j<k}}\left[\alpha V_{ijk}^{\mathrm{R}}+\beta V_{ijk}^{\mathrm{2\pi}}\right] \rangle_{\mathrm{F}}+\gamma\rho^2e^{-\delta\rho}\left[1-(1-2x)^2\right]  \label{eq:E3}. 
\end{equation}
Here, the brackets with the subscript F represent the expectation value with the degenerate Fermi-gas wave function.  
The parameters $\alpha$ and $\beta$ represent the effects of the dynamical correlations among nucleons and the relativistic correction \cite{APR}.  
The last term on the right-hand side of Eq. ({\ref{eq:E3}}) is a phenomenological correction term. 
\footnote{The validity of this treatment of the three-body force is examined for SNM and PNM in Ref. \cite{TBF}.}
This $E_3/N$ for ANM is a straightforward extension of those for SNM and PNM given in papers I and II, as explained below.  

The first term on the right-hand side of Eq. (\ref{eq:E3}) depends on $x$, due to the $x$-dependence of the degenerate Fermi-gas wave function for ANM.  
Since we have chosen in papers I and II to treat $\alpha$ and $\beta$ as independent of $x$ to obtain reasonable values of $E/N$ for both SNM and PNM, we make the same assumption for these parameters in this paper.  
In contrast, in paper I, we introduced a correction term corresponding to the second term on the right-hand side of Eq. (\ref{eq:E3}) in the case of SNM only, and set $\gamma=0$ for PNM, following the treatment by APR.  
In this paper, therefore, we simply assume that the correction term goes to zero quadratically as $x$ approaches zero, as explicitly shown in the last factor in Eq. (\ref{eq:E3}):
The $x$-dependence of the energy of ANM will be discussed in more detail later.  

Finally, the total energy per nucleon $E/N$ is expressed as
\begin{equation}
\frac{E}{N}=\frac{E_2}{N}+\frac{E_3}{N}.
\end{equation}
The parameters $\alpha$, $\beta$, $\gamma$ and $\delta$ in Eq. (\ref{eq:E3}) were determined in paper I so that the obtained $E/N$ could reproduce the empirical saturation density $\rho_0$, saturation energy $E_0/N$, incompressibility $K$ and symmetry energy $E_{\mathrm{sym}}\equiv E(\rho_0,x=0)/N-E(\rho_0,x=1/2)/N$. 
Furthermore, in paper II, the values of these parameters were fine-tuned so that the Thomas-Fermi calculations for isolated atomic nuclei with the obtained $E/N$ could reproduce the gross features of the experimental data in their masses and radii:
For our purpose of constructing a reliable SN-EOS, this fine-tuning is an important process.  
Explicitly, $\alpha=0.43$, $\beta=-0.34$, $\gamma=-1804 \mathrm{MeV fm^6}$ and $\delta=14.62 \mathrm{fm^3}$ to obtain $\rho_0 = 0.16 \mathrm{fm^{-3}}$, $E_0/N = -16.09$ MeV, $K = 245$ MeV and $E_{\mathrm{sym}} = 30.0$ MeV.  
\footnote{We note that, due to the improvements in the numerical calculations, the values of the physical quantities shown in this paper are slightly different from those previously given in papers I and II .  }

\begin{figure}
  \centering
  \includegraphics[width=10.0cm]{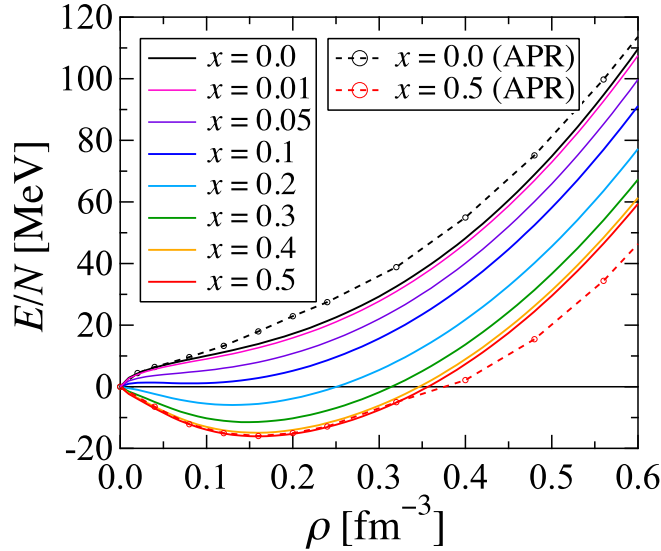}
  \caption{Energies per nucleon, $E/N$, for various values of the proton fraction $x$ as functions of the number density $\rho$. The energies obtained by APR for SNM and PNM are also shown.}
\label{fig:E0}
\end{figure}

The obtained $E/N$ for ANM are shown in Fig. \ref{fig:E0}: Also shown are the results of APR.  
In the case of SNM ($x=0.5$) at $\rho \lesssim 0.32 \mathrm{fm}^{-3}$, the present $E/N$ is in good agreement with that obtained by APR.  
At $\rho \gtrsim 0.32 \mathrm{fm}^{-3}$, however, our $E/N$ becomes higher than that obtained by APR, due to the $\pi^0$ condensation occurring in the study by APR.  
As $x$ decreases, $E/N$ increases monotonically.  
Then, in the case of PNM at $x=0$, the present $E/N$ is lower than that obtained by APR except for in the low-density region, because the symmetry energy obtained by APR is about 34 MeV, which is larger than the present result.  

Next, we apply the obtained $E(\rho,x)/N$ to neutron stars (NSs).  
As in papers I and II, we treat NS matter as a charge-neutral, $\beta$-stable mixture of nucleons, electrons and muons at zero temperature.  
For the NS crust, we employ the crust EOS calculated by the Thomas-Fermi method in paper II. 
We note that the obtained EOS of NS matter violates the causality, i.e., the sound velocity exceeds the speed of light, 
at densities higher than the critical density $\rho_{\mathrm{c}}$ = 0.88 $\mathrm{fm}^{-3}$, which is close to the corresponding value of the EOS by APR \cite{APR}.  
In this paper, therefore, we construct a modified causal EOS by following the method of Ref. \cite{Rhoades}, i.e., at densities higher than $\rho_{\mathrm{c}}$, 
we replace the pressure by $P(\varepsilon)=P(\varepsilon_{\mathrm{c}})+\varepsilon-\varepsilon_{\mathrm{c}}$, where $\varepsilon$ is the energy density and $\varepsilon_{\mathrm{c}}$ is the energy density at $\rho_{\mathrm{c}}$.  
  
\begin{figure}[t]
  \centering
  \includegraphics[width=10.0cm]{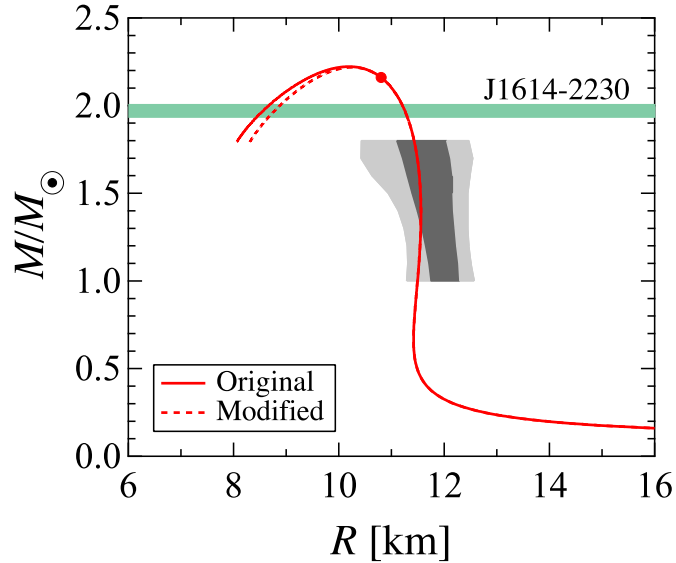}
  \caption{Mass-radius relation of NSs with the present EOS. 
  The filled circle represents the NS for which the central density is equal to the critical density $\rho_{\mathrm{c}}$:   
  For smaller radii, the solid and dotted lines show the masses of NSs with the original EOS and the modified causal one, respectively.  
The horizontal band shows the mass of PSR J1614-2230 ($1.97 \pm 0.04 M_{\odot}$) \cite{2Msolar}.
The shaded region is given in Ref. \cite{Steiner} based on the observational data: The dark and light gray regions correspond to 1$\sigma$ and 2$\sigma$ errors, respectively. 
}
\label{fig:NS}
\end{figure}

With the original and modified EOSs, we solve the TOV equation to obtain the masses and radii of NSs.  
Figure \ref{fig:NS} shows the obtained mass-radius relation of NSs.  
It can be seen that the stable NS solutions with the original and modified EOSs are hardly distinguishable from each other, and both the maximum masses are 2.22$M_\odot$.  
Also shown in this figure are the mass of the heavy NS J1614-2230 ($M=1.97 \pm 0.04 M_\odot$) \cite{2Msolar} and the observationally suggested region analyzed in Ref. \cite{Steiner}.  
The masses and radii of NSs with the present EOS are consistent with those of the observational data.  

\subsection{Proton-fraction-dependence of the symmetry energy}
In this subsection, we discuss the proton-fraction-dependence of $E/N$ and the symmetry energy $E_\mathrm{sym}$ in more detail. 
For this purpose, we define the following energy difference:
\begin{equation}
E_{\mathrm{I}}(\rho,x) \equiv \frac{E(\rho,x)}{N}-\frac{E(\rho,x=1/2)}{N}.    \label{eq:sym1}
\end{equation}
From this definition, it is obvious that $E_\mathrm{sym}=E_\mathrm{I}(\rho_0,0)$.  
In many cases, this symmetry energy $E_\mathrm{sym}$ is regarded as the same as $S_\mathrm{I}(\rho_0)$ given by the original definition, i.e., 
\begin{equation}
S_\mathrm{I}(\rho) \equiv \frac{1}{8}\frac{\partial^2}{\partial x^2}\frac{E(\rho, x)}{N}\Biggr |_{x=1/2}.  \label{eq:sym2}
\end{equation}
These two quantities, $E_\mathrm{sym}$ and $S_\mathrm{I}(\rho_0)$, are identical if, as is usually assumed, $E(\rho,x)/N$ increases quadratically with $x$ from $x=1/2$ (SNM) to $x=0$ (PNM), or equivalently, if $E_{\mathrm{I}}(\rho,x)$ is proportional to $\zeta \equiv (1-2x)^2$. 

\begin{figure}[t]
  \centering
  \includegraphics[width=10.0cm]{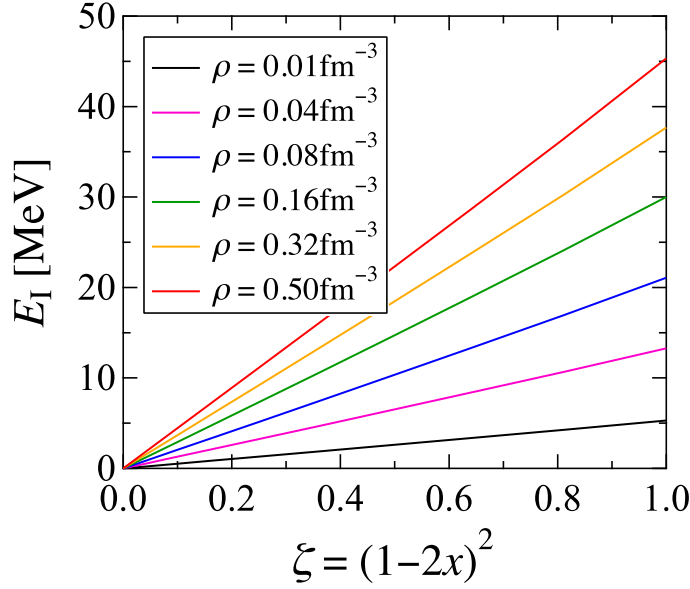}
  \caption{Energy differences $E_{\mathrm{I}}(\rho,x)$ given in Eq. (\ref{eq:sym1}) for various values of $\rho$ as functions of $\zeta$.}
\label{fig:Esym1}
\end{figure}

In order to examine this $x$-dependence of $E_{\mathrm{I}}(\rho,x)$ in the present calculation, we show $E_{\mathrm{I}}$ at various densities as functions of $\zeta$ in Fig. \ref{fig:Esym1}.
It can be seen that, in the wide range of densities, $E_{\mathrm{I}}$ is nearly proportional to $\zeta$.  

\begin{figure}[t]
  \centering
  \includegraphics[width=14.0cm]{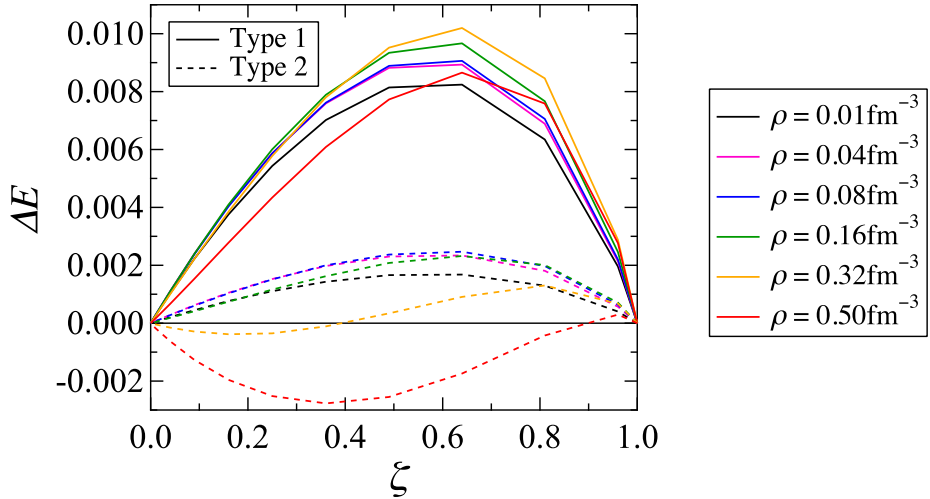}
  \caption{Relative deviations $\mathnormal{\Delta} E$ of the interpolated energies from the original ones as functions of $\zeta$ (Type 1). 
  The dotted curves show $\mathnormal{\Delta} E$ with the interpolation by Lagaris and Pandharipande, which is discussed in Appendix B (Type 2).  }
\label{fig:delta0}
\end{figure}

For a more quantitative examination, we try to evaluate $E(\rho,x)/N$ by interpolating the energy per nucleon for PNM, $E(\rho, 0)/N$, and that for SNM, $E(\rho, 1/2)/N$, as
\begin{equation}
\frac{E_{\mathrm{Int}}(\rho, x)}{N}=\frac{E(\rho, 1/2)}{N}+\Bigg[\frac{E(\rho, 0)}{N}-\frac{E(\rho, 1/2)}{N}\Bigg](1-2x)^2.    \label{eq:Type1}
\end{equation}
Then we show in Fig. \ref{fig:delta0} the relative deviation of $E_{\mathrm{Int}}(\rho,x)/N$ from $E(\rho,x)/N$, defined by
\begin{equation}
\mathnormal{\Delta} E(\rho, x) \equiv \frac{E_{\mathrm{Int}}(\rho, x)/N - E(\rho, x)/N}{E_{\mathrm{I}}(\rho,x)}.   \label{eq:deltaE}
\end{equation}
It can be seen that $\mathnormal{\Delta} E(\rho, x)$ is concave with respect to $\zeta$, and $0 \leq \mathnormal{\Delta} E \lesssim 0.01$ at the densities shown in this figure.  
The main cause of this concaveness is the one-body kinetic energy $E_1/N$ which is a convex function of $\zeta$.  
The dotted curves in Fig. \ref{fig:delta0} will be explained in Appendix B.  

\begin{figure}[t]
  \centering
  \includegraphics[width=10.0cm]{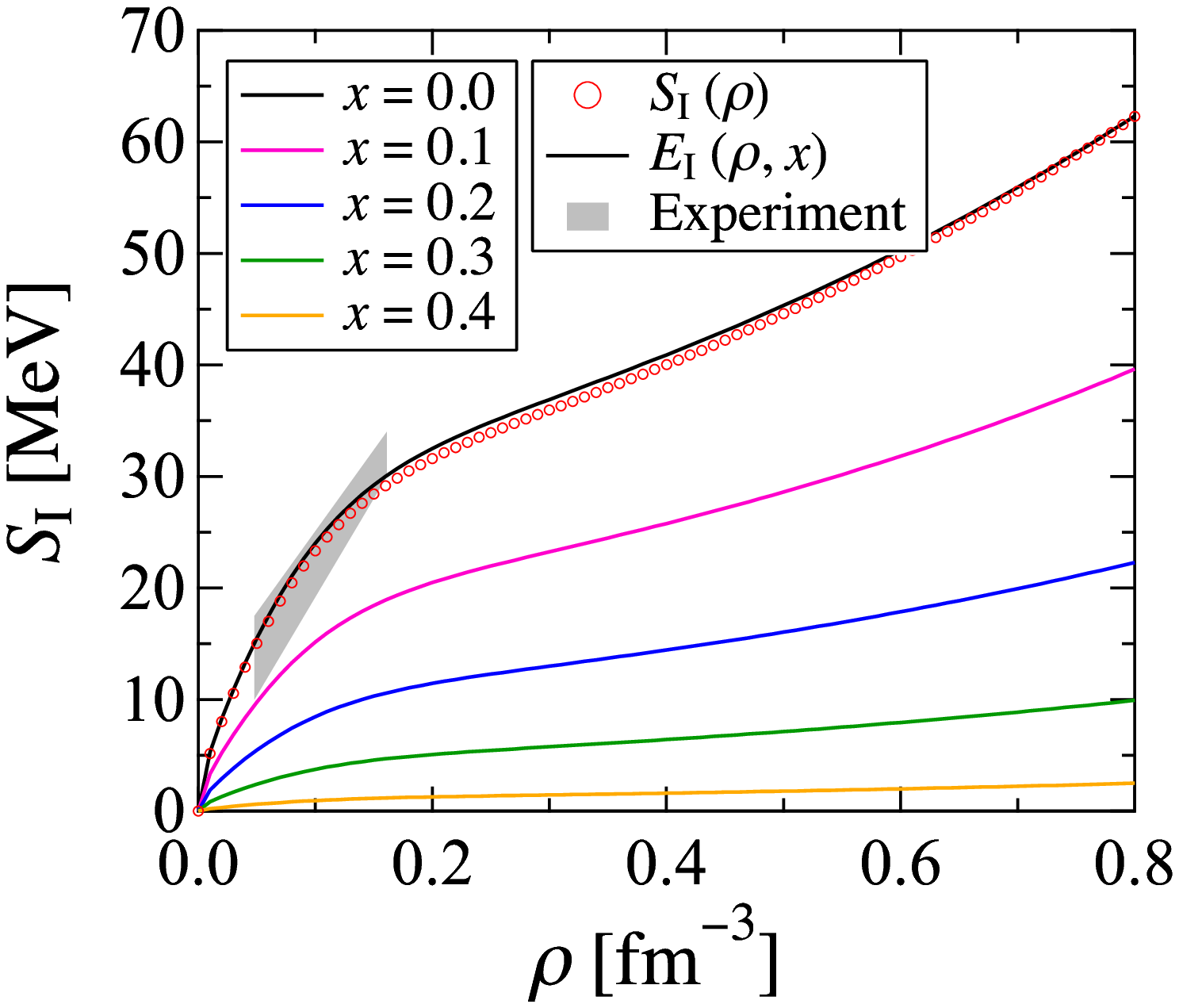}
  \caption{The density-dependent symmetry energy $S_\mathrm{I}(\rho)$ defined by Eq. (\ref{eq:sym2}) as a function of the density $\rho$.  
  $E_{\mathrm{I}}(\rho,x)$ are also shown for various values of $x$. 
                  The shaded region shows the experimental data on $S_\mathrm{I}(\rho)$ \cite{expS}.}
\label{fig:Esym2}
\end{figure}

Figure \ref{fig:Esym2} shows the density dependence of $S_\mathrm{I}(\rho)$ and $E_\mathrm{I}(\rho,x)$.
As can be seen in Fig. \ref{fig:Esym2}, $S_\mathrm{I}(\rho)$ is slightly lower than $E_{\mathrm{I}}(\rho,0)$, because, in Fig. \ref{fig:Esym1}, $E_{\mathrm{I}}(\rho,x)$ is slightly convex with respect to $\zeta$, 
and $S_\mathrm{I}(\rho)$ corresponds to the gradient of $E_{\mathrm{I}}(\rho,x)$ at $\zeta = 0$.  
The shaded region in Fig. \ref{fig:Esym2} shows the experimental data on $S_\mathrm{I}(\rho)$ obtained from heavy-ion collisions \cite{expS}; the symmetry energy of the present EOS is consistent with those experimental data. 

\section{Asymmetric nuclear matter at finite temperatures}
\subsection{Extension of the variational method by Schmidt and Pandharipande}
In this subsection, we calculate the free energy per nucleon $F/N$ of ANM as a function of the number density $\rho$, proton fraction $x$ and temperature $T$.
To obtain $F/N$ for SNM and PNM, in paper I, we adopted the variational method by Schmidt and Pandharipande (SP) \cite{SP}. 
In this method, $F/N$ is expressed with the effective mass of a nucleon $m^*$, and then $F/N$ is minimized with respect to $m^*$, as explained below in more detail.  
The validity of this method was examined in Ref. \cite{MP}, and has been employed in some many-body calculations of hot SNM and PNM \cite{AM,Illarionov,FP}.
Since our purpose is to treat ANM, we extend the method by SP to ANM, i.e., we distinguish the effective masses of a proton and a neutron $m_i^*$ ($i=$p, n), and treat them independently.  

We start from the following expression: 
\begin{equation}
\frac{F}{N} = \frac{E_{\mathrm{0T}}}{N}-T\frac{S_0}{N},    \label{eq:FE}
\end{equation}
where $E_{\mathrm{0T}}/N$ is the approximate internal energy per nucleon and $S_0/N$ is the approximate entropy per nucleon at temperature $T$. 
Here, the adjective "approximate" is attached in order to distinguish these quantities from those calculated with $F/N$ through the thermodynamic relations.  

As in the case of zero temperature, we express $E_\mathrm{0T}/N$ as a sum of the two-body internal energy $E_{\mathrm{2T}}/N$ and the three-body internal energy $E_{\mathrm{3T}}/N$:
\begin{equation}
\frac{E_{\mathrm{0T}}}{N}=\frac{E_{\mathrm{2T}}}{N}+\frac{E_{\mathrm{3T}}}{N}.          \label{eq:E0T}
\end{equation}
At zero temperature, $E_\mathrm{2T}/N$ reduces to $E_2/N$ given in Eq. (\ref{eq:E2}), where the occupation probabilities of single-particle states in the Jastrow wave function Eq. (\ref{eq:Jastrow}) are expressed as $n_{0i}(k)=\theta(k_{\mathrm{F}i}-k)$ $(i=\mathrm{p}, \mathrm{n})$.   
At finite temperatures, we follow the method of SP, and construct $E_{\mathrm{2T}}/N$ by replacing $n_{0i}(k)$ in $E_2/N$ at zero temperature by the following averaged occupation probabilities $n_{i}(k)$:
\begin{equation}
n_i(k) = \Bigg\{1+\exp \left[\frac{\varepsilon_i(k)-\mu_{0i}}{k_{\mathrm{B}}T}\right]\Bigg\}^{-1},    \label{eq:nk}
\end{equation}
where $\varepsilon_i(k)$ are the quasi-nucleon energies which are assumed to be as follows:
\begin{equation}
\varepsilon_i(k) = \frac{\hbar^2k^2}{2m^\ast_i}.         \label{eq:epsilon}
\end{equation}
Here, $m^\ast_i$ $(i = \mathrm{p}, \mathrm{n})$ are the effective masses for protons and neutrons, respectively.   
In Eq. (\ref{eq:nk}), the values of $\mu_{0i}$ are determined by the following normalization conditions:
\begin{equation}
\rho_i = \frac{1}{\pi^2}\int_0^{\infty}n_i(k)k^2dk,      \label{eq:mu0}
\end{equation}
where $\rho_\mathrm{p}$ and $\rho_\mathrm{n}$ are the proton and neutron number densities, respectively.  
The explicit expression of $E_{\mathrm{2T}}/N$ obtained in this method is similar to $E_2/N$ at zero temperature given by Eq. (\ref{eq:E2}), but $E_1/N$, $F_{\mathrm{F}ts}^{\mu}(r)$, $F_{\mathrm{qF}ts}^{\mu}(r)$ and $F_{\mathrm{bF}ts}^{\mu}(r)$ are replaced by $E_{\mathrm{1T}}/N$, $F_{\mathrm{F}ts}^{\mu}(r; T)$, $F_{\mathrm{qF}ts}^{\mu}(r; T)$ and $F_{\mathrm{bF}ts}^{\mu}(r; T)$, the explicit expressions of which are given in Appendix A. 
Following the method by SP, the two-body correlation functions $f^{\mu}_{\mathrm{C}ts}(r)$, $f^{\mu}_{\mathrm{T}t}(r)$ and $f^{\mu}_{\mathrm{SO}t}(r)$ are assumed to be the same as at zero temperature: 
We refer to this approximation as the frozen-correlation approximation (FCA), the validity of which will be discussed later.  

The three-body internal energy $E_{\mathrm{3T}}/N$ in Eq. (\ref{eq:E0T}) is assumed to be the same as $E_3/N$ at zero temperature in Eq. (\ref{eq:E3}), as in paper I, where the validity of this assumption is also discussed \cite{K1}. 

In the method of SP, the approximate entropy $S_0/N$ in Eq. (\ref{eq:FE}) is expressed as in the case of a gas of quasi-nucleons whose masses are the effective mass $m^*$. 
In this paper, we extend the method so as to treat ANM, i.e., 
\begin{equation}
\frac{S_0}{N} = -\sum_{i=\mathrm{p,n}} \frac{k_{\mathrm{B}}}{\pi^2\rho}\int_0^{\infty}
\Bigl\{\left[1-n_i(k)\right]\ln\left[1-n_i(k)\right]+n_i(k)\ln n_i(k)\Bigl\}k^2dk.
\end{equation}
Then, $F/N$ is expressed as an explicit function of $m^\ast_{\mathrm{p}}$ and $m^\ast_{\mathrm{n}}$, and is minimized with respect to them.
\subsection{Free energy and related thermodynamic quantities}

\begin{figure}
\begin{center}
    \includegraphics[width=10cm]{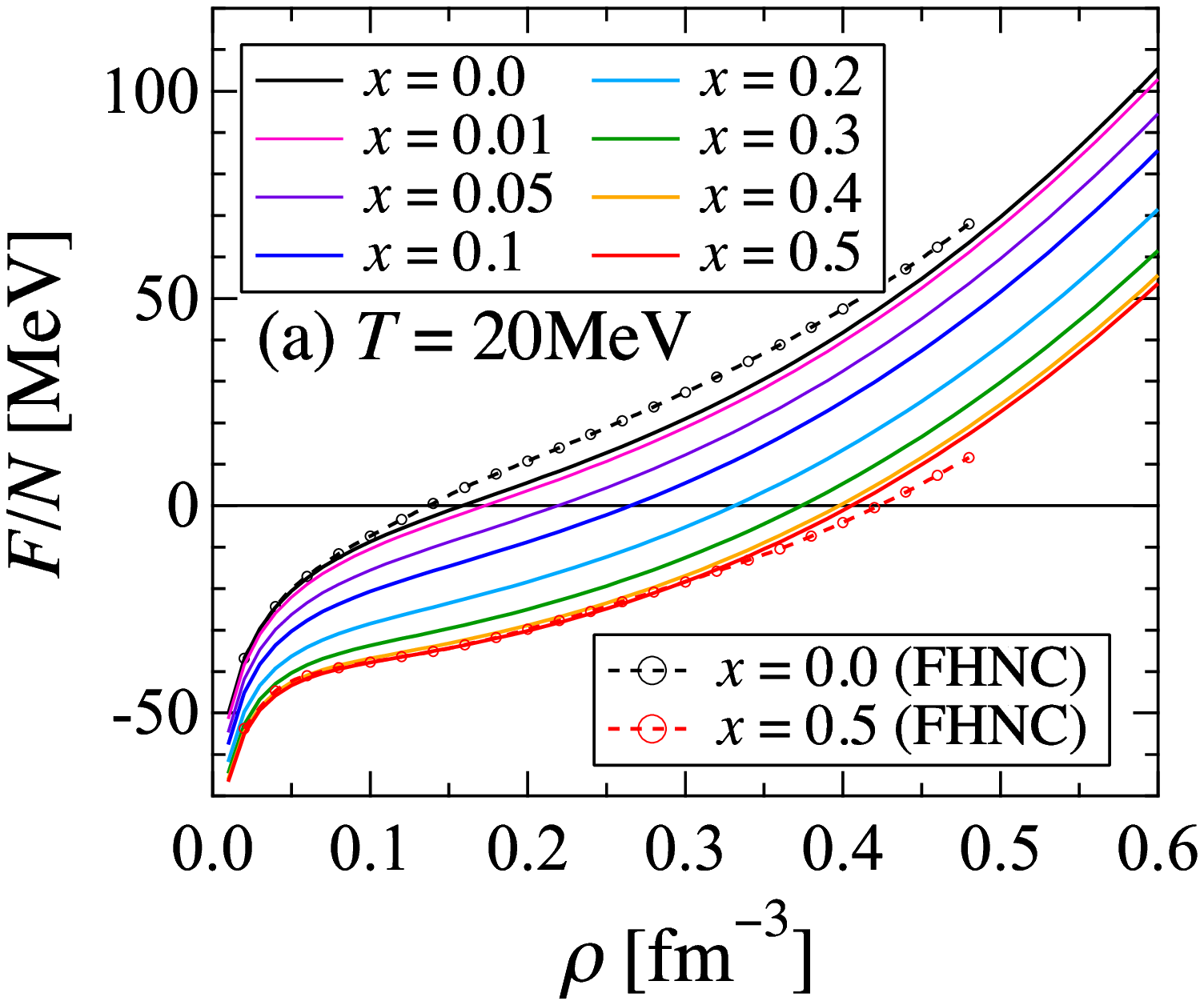}
\end{center}
\begin{center}
    \includegraphics[width=10cm]{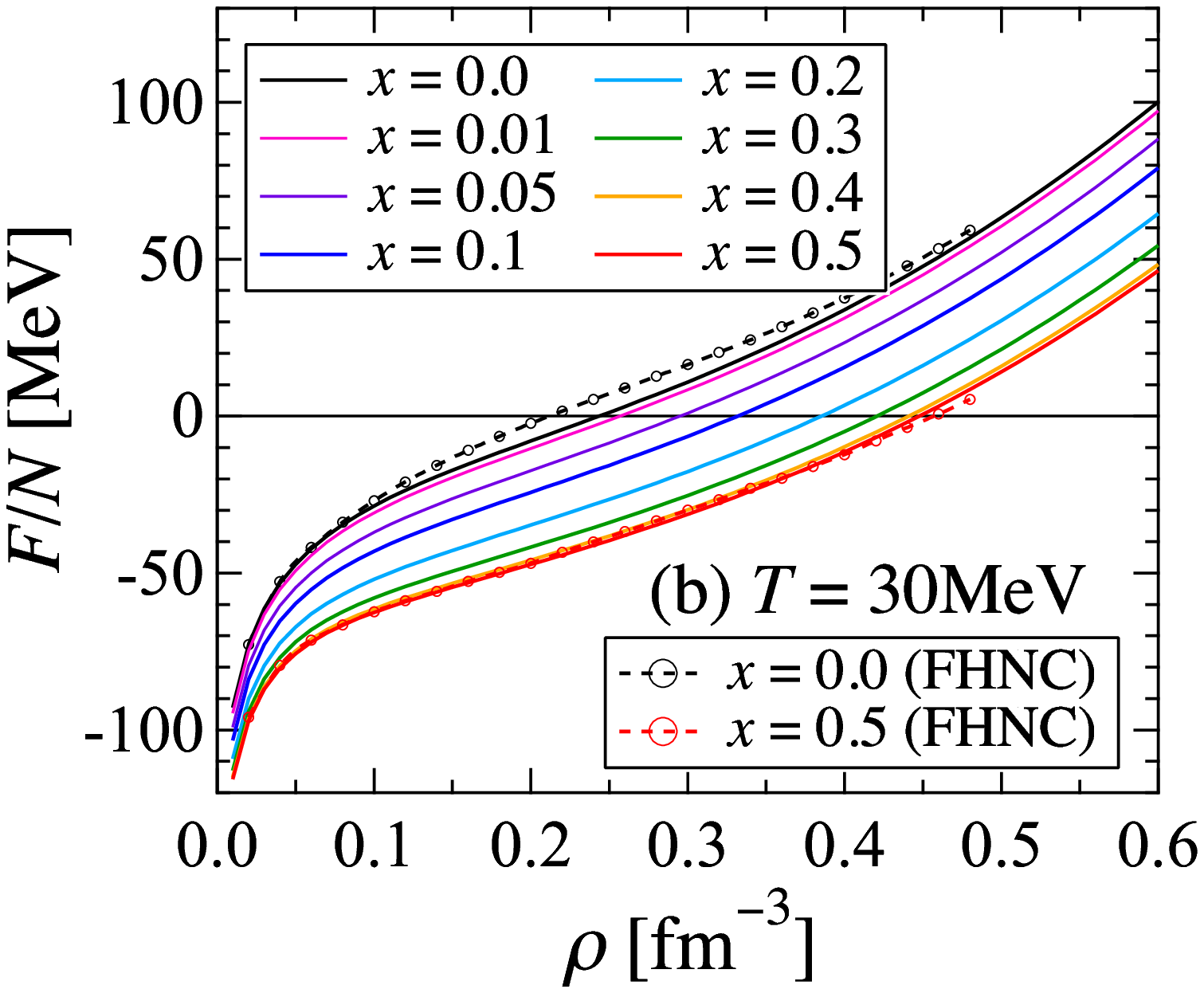}
\end{center}
\caption{Free energies per nucleon at $T = 20$ MeV (a) and $T = 30$ MeV (b) for various values of $x$ as functions of $\rho$.
The free energies calculated with the FHNC method for SNM and PNM given in Ref. \cite{AM} are also shown.}
\label{fig:F}
\end{figure}

The obtained free energies per nucleon of ANM are shown in Fig. \ref{fig:F}: 
Also shown are those obtained with the FHNC calculations for SNM and PNM given in Ref. \cite{AM}, which are extensions of the results of APR.  
As in the case at zero temperature, the present result is in good agreement with that obtained with the FHNC method for SNM.  
As $x$ decreases, $F/N$ increases monotonically, and, in the case of PNM, our result is somewhat lower than that calculated with the FHNC method because of the larger symmetry energy in the case of APR, as pointed out in the last section at zero temperature.  

\begin{figure}[t]
\begin{center}
    \includegraphics[width=10cm]{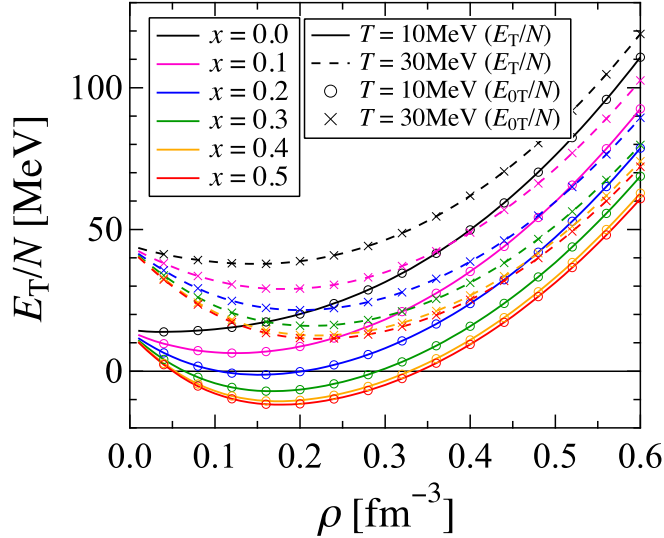}
\end{center}
\caption{Internal energies per nucleon, $E_{\mathrm{T}}/N$, at $T=$10 and 30 MeV for various values of $x$ as functions of $\rho$. 
The approximate internal energies per nucleon, $E_{\mathrm{0T}}/N$, are also shown.}
\label{fig:UU}
\end{figure}

Figure \ref{fig:UU} shows the internal energy per nucleon $E_{\mathrm{T}}/N$ of ANM derived from $F/N$ through the thermodynamic relation.  
$E_{\mathrm{T}}/N$ increases as $x$ decreases monotonically from SNM to PNM.  
Also shown is the approximate internal energy $E_{\mathrm{0T}}/N$. 
It can be seen that $E_{\mathrm{T}}/N$ is in good agreement with $E_{\mathrm{0T}}/N$ for various values of $x$, which implies that the present variational method is self-consistent at finite temperatures.  
In paper I, we confirmed this self-consistency for SNM and PNM: 
In this paper, we have confirmed that this is also true for ANM.  

\begin{figure}[t]
\begin{center}
    \includegraphics[width=10cm]{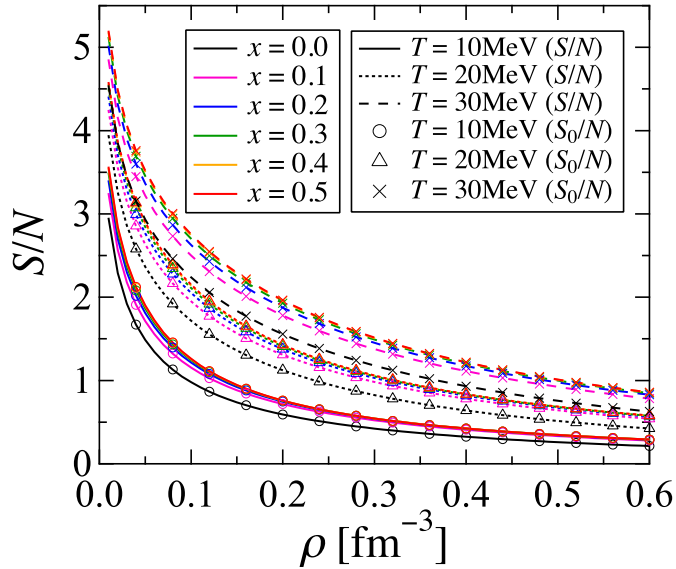}
\end{center}
\caption{Entropies per nucleon, $S/N$, at $T=$10, 20 and 30 MeV for various values of $x$ as functions of $\rho$. 
The approximate entropies per nucleon, $S_{\mathrm{0}}/N$, are also shown.}
\label{fig:SS}
\end{figure}

Figure \ref{fig:SS} shows the entropy per nucleon $S/N$ of ANM derived from $F/N$: 
It can be seen that $S/N$ increases with $x$ monotonically.  
The approximate entropy per nucleon $S_0/N$ is also shown: $S/N$ and $S_0/N$ are in good agreement with each other, too.  

\begin{figure}[t]
\begin{center}
    \includegraphics[width=10cm]{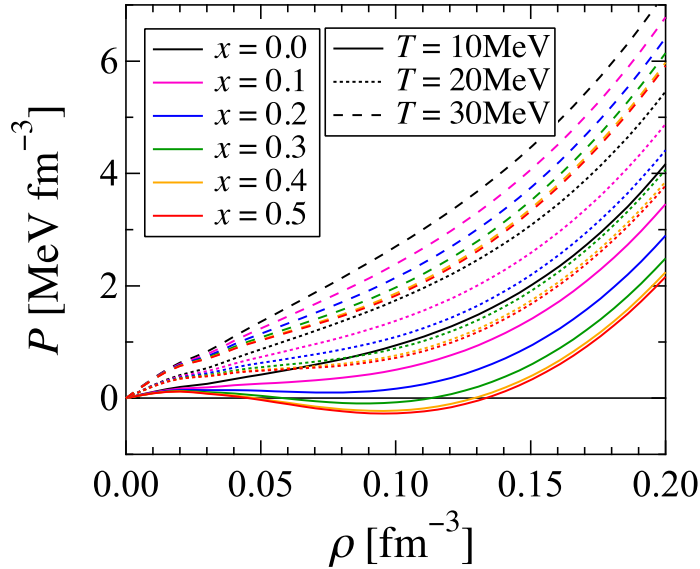}
\end{center}
\caption{Pressures at $T=$10, 20 and 30 MeV for various values of $x$ as functions of $\rho$. }
\label{fig:P}
\end{figure}

The pressure $P$ of ANM is shown in Fig. \ref{fig:P}: 
It can be seen that $P$ increases monotonically as $x$ decreases.  
For a fixed $x$, $P$ increases with $T$, and at $T$ higher than a critical value $T_{\mathrm{0}}$, $P$ increases monotonically with $\rho$, i.e., at $T < T_{\mathrm{0}}$, the mechanical instability occurs.  
In the case of SNM, $T_{\mathrm{0}}$ is equal to the critical temperature $T_{\mathrm{c}}$ of the liquid-gas phase transition \cite{MS}, and $T_{\mathrm{0}} \simeq 18$ MeV, as reported in paper I.  
It is found that $T_{\mathrm{0}}$ decreases with $x$, and becomes 0 MeV at $x\lesssim $ 0.08. 
Here it should be noted that, in order to determine $T_{\mathrm{c}}$ of ANM, the chemical instability, in addition to the mechanical instability, has to be taken into account, as discussed in Ref. \cite{MS}.  
In our project to construct an SN-EOS, as mentioned before, we will treat this non-uniform mixed phase in the Thomas-Fermi approximation.  

\begin{figure}
  \centering
  \includegraphics[width=10.0cm]{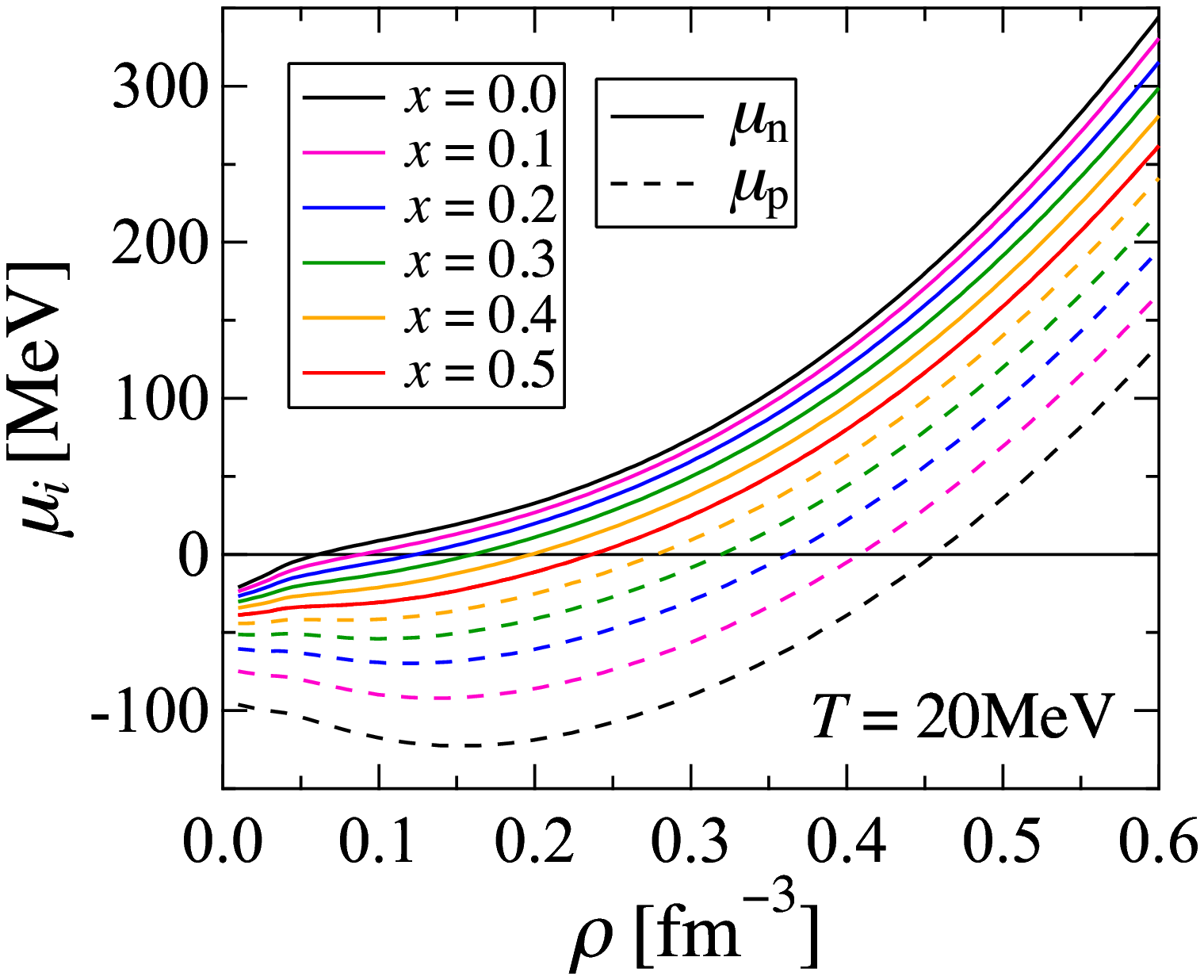}
  \caption{The proton chemical potential $\mu_{\mathrm{p}}$ and the neutron chemical potential $\mu_{\mathrm{n}}$ at $T =$ 20 MeV for various values of $x$ as functions of $\rho$.}
\label{fig:chem}
\end{figure}

Figure \ref{fig:chem} shows the proton chemical potential $\mu_{\mathrm{p}}$ and the neutron chemical potential $\mu_{\mathrm{n}}$ at $T=20$ MeV: 
In the case of SNM, $\mu_{\mathrm{p}}=\mu_{\mathrm{n}}$, and, as $x$ decreases, $\mu_{\mathrm{n}}$ decreases while $\mu_{\mathrm{p}}$ increases.  
It should be noted that $\mu_{\mathrm{p}}$ and $\mu_{\mathrm{n}}$ shown in this figure are derived from $F/N$ with the thermodynamic relations, and are different from $\mu_{0i}$ introduced in Eq. (\ref{eq:nk}): 
The latter include the effect of the single-particle potential energies that are not included in $\varepsilon_i(k)$ given by Eq. (\ref{eq:epsilon}).

\begin{figure}
  \centering
  \includegraphics[width=10.0cm]{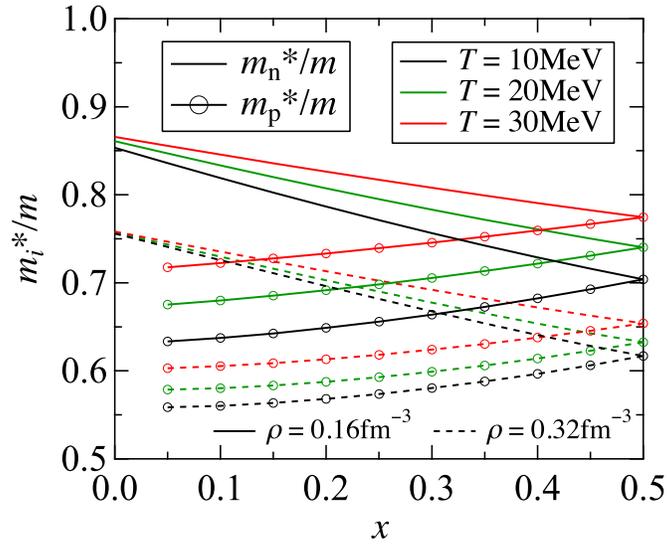}
  \caption{Proton effective masses $m^\ast_{\mathrm{p}}$ and neutron effective masses $m^\ast_{\mathrm{n}}$ for $T=$ 10, 20 and 30 MeV at $\rho =$ 0.16 and 0.32 fm$^{-3}$ as functions of $x$.} 
\label{fig:em}
\end{figure}

The effective masses $m^\ast_i$ are shown in Fig. \ref{fig:em} as functions of $x$. 
In the case of SNM, $m^\ast_{\mathrm{p}}$ is identical to $m^\ast_{\mathrm{n}}$, 
and, as $x$ decreases, $m^\ast_{\mathrm{p}}$ decreases while $m^\ast_{\mathrm{n}}$ increases. 
It is also seen that, as $\rho$ increases, the $T$-dependence of $m^\ast_i$ becomes smaller.  
It should be noted that those $m^\ast_i$ are the variational parameters in the present variational method, but, as pointed out in Ref. \cite{AM}, those can be regarded as weighted averages of the momentum-dependent effective masses of nucleons near the Fermi surfaces.  
In fact, the behavior of the obtained $m^\ast_i$ is fairly reasonable.  

Next, we discuss the validity of the frozen-correlation approximation (FCA) employed in this paper.  
This FCA is adopted in the original paper by SP, and the FHNC calculation in FCA is performed in Ref. \cite{FP}.  
In principle, however, $F/N$ should be minimized with respect to the correlation function $f_{ij}$ as well as $m^*$, as performed in the recent studies \cite{AM,Illarionov}.  
Therefore, in the rest of this subsection, we perform the full minimization of $F/N$, i.e., for each set of $m^*_i$, 
we minimize $F/N$ with respect to $f^{\mu}_{\mathrm{C}ts}(r)$, $f^{\mu}_{\mathrm{T}t}(r)$ and $f^{\mu}_{\mathrm{SO}t}(r)$ 
by solving the Euler-Lagrange (EL) equations derived from Eq. (\ref{eq:FE}), and then minimize $F/N$ with respect to $m^*_i$.  
In solving the EL equations, we impose two constraints as in the case at zero temperature, i.e., the extended Mayer's condition and the healing-distance condition. 
The former is a condition similar to Eq. (\ref{eq:mayer}), but $F_{\mathrm{F}ts}^{\mu}(r)$ and $F_{\mathrm{qF}ts}^{\mu}(r)$ in Eqs. (\ref{eq:Fts}) and (\ref{eq:mayer}) are replaced by $F_{\mathrm{F}ts}^{\mu}(r; T)$ and $F_{\mathrm{qF}ts}^{\mu}(r; T)$, respectively.  
In the latter condition, the healing distance is assumed to be the same as at zero temperature.  

\begin{figure}[t]
  \centering
  \includegraphics[width=10.0cm]{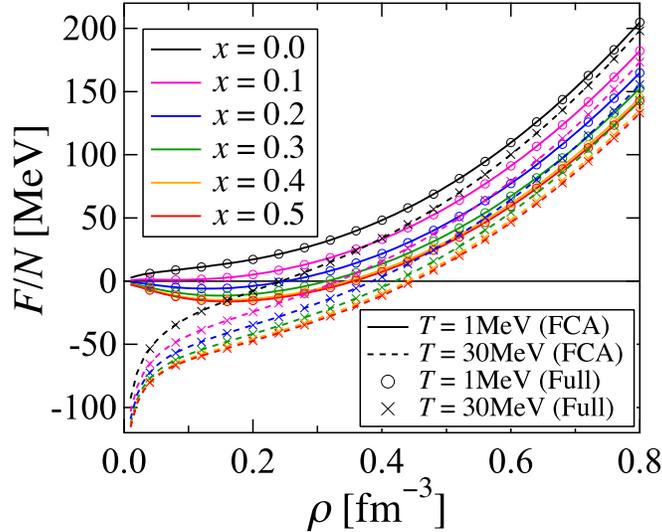}
  \caption{Free energies per nucleon at $T=$ 1 and 30 MeV for various values of $x$ in the frozen-correlation approximation and those with the full minimization as functions of $\rho$.}
\label{fig:Fully}
\end{figure}

In Fig. \ref{fig:Fully}, the fully minimized values of $F/N$ are compared with those of the FCA. 
It is found that the values of $F/N$ in the FCA are in good agreement with those with the full minimization in a wide range of $\rho$, $x$ and $T$.  
This agreement implies that the effects of the full minimization and the extended Mayer's condition cancel out each other.  
As a result, it is confirmed that the FCA is a good approximation in the present calculation, which is consistent with the result in Ref. \cite{FP}.  

\subsection{Proton-fraction-dependence of the free energy}
In this subsection, we discuss the $x$-dependence of $F/N$.  
First, we define the following free energy difference: 
\begin{equation}
F_{\mathrm{I}}(\rho, x, T)\equiv \frac{F(\rho, x, T)}{N}-\frac{F(\rho, x=1/2, T)}{N},    \label{eq:Fsym1}
\end{equation}
and refer to $F_\mathrm{sym}(\rho,T)\equiv F_\mathrm{I}(\rho,x=0,T)$ as the symmetry free energy.  
At zero temperature, $F_\mathrm{sym}(\rho,T)$ reduces to the symmetry energy; $E_\mathrm{sym} = F_\mathrm{sym}$ ($\rho_0$, $T$=0 MeV).  
\begin{figure}[t]
  \centering
  \includegraphics[width=10.0cm]{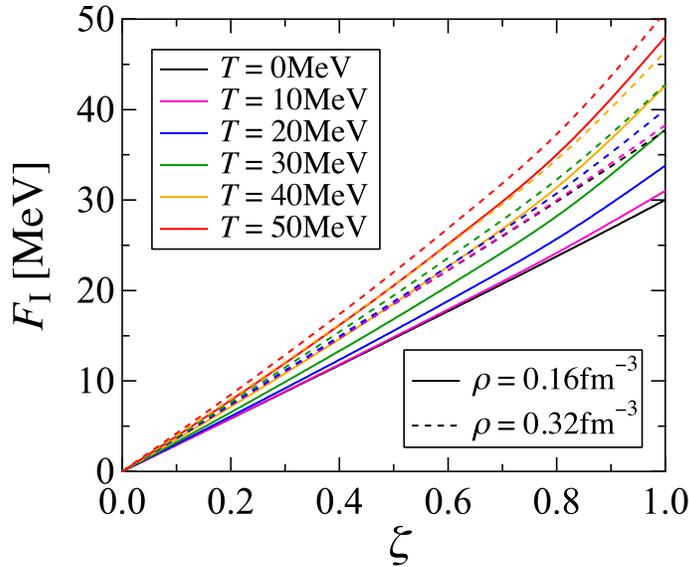}
  \caption{Free energy differences $F_{\mathrm{I}}/N$ given in Eq. (\ref{eq:Fsym1}) for various values of $T$ at $\rho = 0.16$ and 0.32 $\mathrm{fm}^{-3}$ as functions of $\zeta$.}
\label{fig:fsym}
\end{figure}

Figure \ref{fig:fsym} shows $F_{\mathrm{I}}(\rho,x, T)$ as functions of $\zeta = (1-2x)^2$. 
Contrary to the case at zero temperature, it can be seen, in this figure, that $F_{\mathrm{I}}(\rho,x, T)$ are not proportional to $\zeta$: 
The curves deviate from the straight lines more markedly as the temperature increases.  

\begin{figure}
  \centering
  \includegraphics[width=14.0cm]{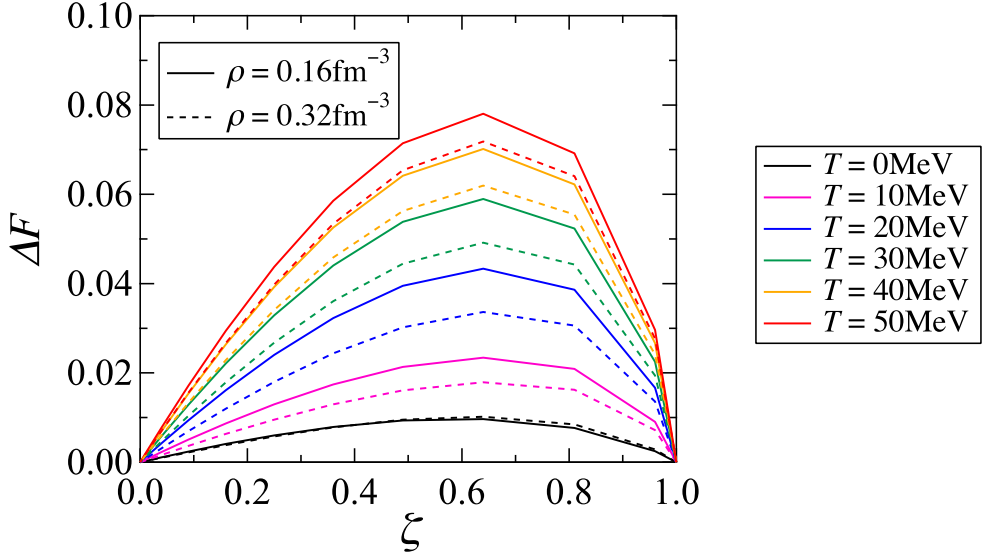}
  \caption{Relative deviations $\mathnormal{\Delta} F$ of the interpolated free energies from the original ones as functions of $\zeta$.}
\label{fig:deltaF}
\end{figure}
For a more quantitative discussion, as in the case at zero temperature, we evaluate $F(\rho,x, T)/N$ by a linear interpolation between $F/N$ of SNM and that of PNM with respect to $\zeta$, i.e., 
\begin{equation}
\frac{F_{\mathrm{Int}}(\rho, x, T)}{N}=\frac{F(\rho, 1/2, T)}{N}+\Bigg[\frac{F(\rho, 0, T)}{N}-\frac{F(\rho, 1/2, T)}{N}\Bigg](1-2x)^2, 
\end{equation}
and then calculate the relative deviation $\mathnormal{\Delta} F(\rho,x,T)$ defined by
\begin{equation}
\mathnormal{\Delta} F(\rho, x, T) = \frac{F_{\mathrm{Int}}(\rho, x, T)/N - F(\rho, x, T)/N}{F_{\mathrm{I}}(\rho, x, T)}.  
\end{equation}
Figure \ref{fig:deltaF} shows $\mathnormal{\Delta} F(\rho,x,T)$ as functions of $\zeta$. 
As seen in this figure, the relative deviations at finite temperatures are much larger than those at zero temperature. 
This implies that the quadratic interpolation between $F/N$ of SNM and PNM with respect to $x$ to obtain that of ANM, which is fairly good at zero temperature, is inappropriate at finite temperatures, as pointed out in Ref. \cite{Illarionov}.

\begin{figure}[t]
  \centering
  \includegraphics[width=10.0cm]{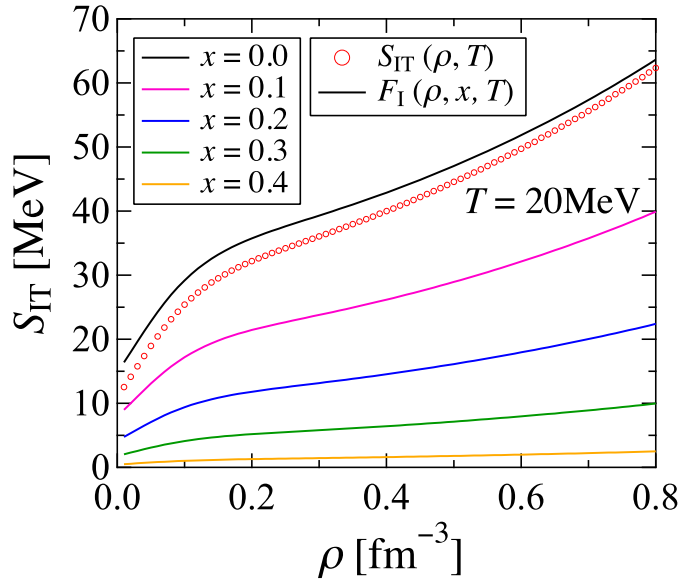}
 \caption{The density-dependent symmetry free energy $S_{\mathrm{IT}}(\rho, T)$ defined by Eq. (\ref{eq:fsym2}) at $T$ = 20MeV as a function of the density $\rho$.  
  $F_{\mathrm{I}}(\rho, x, T)$ at $T$ = 20MeV are also shown for various values of $x$.}
\label{fig:fsym2}
\end{figure}
In order to discuss the density dependence of the deviation mentioned above, we define the density-dependent symmetry free energy $S_{\mathrm{IT}}(\rho, T)$ as
\begin{equation}
S_{\mathrm{IT}}(\rho, T) \equiv \frac{1}{8}\frac{\partial^2}{\partial x^2}\frac{F(\rho, x, T)}{N}\Biggr |_{x=1/2}.  \label{eq:fsym2}
\end{equation}
It is obvious that $S_{\mathrm{IT}}(\rho, T$ = 0MeV$) = S_{\mathrm{I}}(\rho)$ at zero temperature.  
In Fig. \ref{fig:fsym2}, we show $S_{\mathrm{IT}}(\rho, T)$ as well as $F_{\mathrm{I}}(\rho, x, T)$ at $T$ = 20MeV as a function of $\rho$.  
It can be seen that, in the wide range of $\rho$, $S_{\mathrm{IT}}(\rho, T)$ is smaller than $F_{\mathrm{sym}}(\rho, T)=F_{\mathrm{I}}(\rho, x=0, T)$ as in the case at zero temperature, but the difference is much larger, because the curvature of $F_{\mathrm{I}}(\rho, x, T)$ with respect to $\zeta$ in Fig. \ref{fig:fsym} is much larger than that of $E_{\mathrm{I}}(\rho, x)$ shown in Fig. \ref{fig:Esym1}. 
Furthermore, since the thermal effect becomes smaller as $\rho$ increases for a fixed $T$, the difference between $S_{\mathrm{IT}}(\rho, T)$ and $F_{\mathrm{sym}}(\rho, T)$ decreases at high densities.

\begin{figure}[t]
\begin{center}
    \includegraphics[width=10cm]{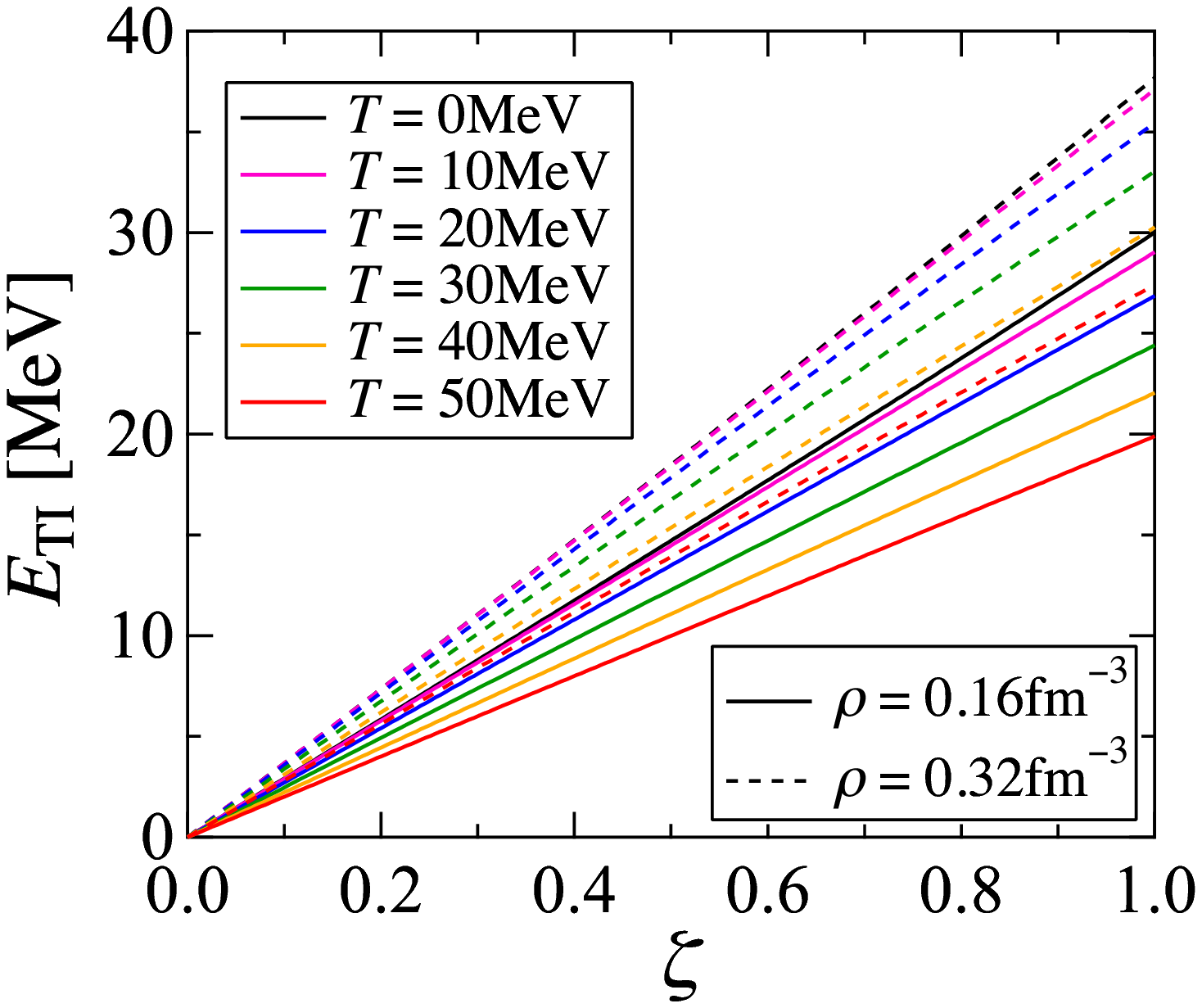}
\end{center}
\caption{Internal energy differences $E_{\mathrm{TI}}$ given in Eq. (\ref{eq:ETI}) for various values of $T$ at $\rho = 0.16$ and 0.32 $\mathrm{fm}^{-3}$ as functions of $\zeta$.}
\label{fig:Usym}
\end{figure}

\begin{figure}[t]
\begin{center}
    \includegraphics[width=10cm]{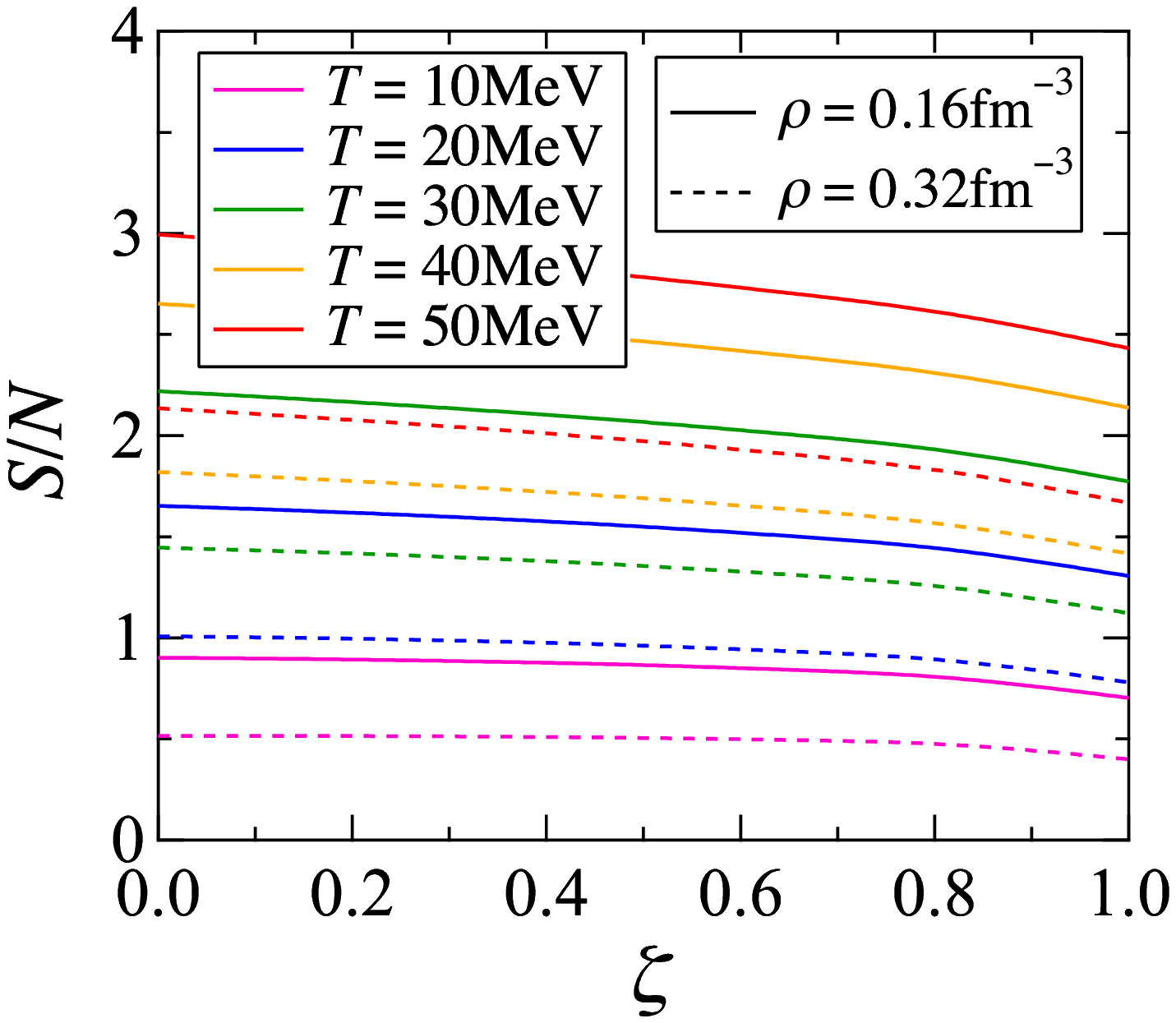}
\end{center}
\caption{Entropies per nucleon $S/N$ for various values of $T$ at $\rho = 0.16$ and 0.32 $\mathrm{fm}^{-3}$ as functions of $\zeta$.}
\label{fig:Ssym}
\end{figure}
Finally, in order to clarify the cause of the relatively-large deviation $\mathnormal{\Delta} F$, we show, as functions of $\zeta$, the following internal energy difference
\begin{equation}
E_{\mathrm{TI}}(\rho, x, T)\equiv \frac{E_{\mathrm{T}}(\rho, x, T)}{N}-\frac{E_{\mathrm{T}}(\rho, 1/2, T)}{N},    \label{eq:ETI}
\end{equation}
and the entropy per nucleon $S(\rho,x,T)/N$, in Figs. \ref{fig:Usym} and \ref{fig:Ssym}, respectively. 
As seen in Fig. \ref{fig:Usym}, $E_{\mathrm{TI}}(\rho,x,T)$ is roughly proportional to $\zeta$, while $S(\rho,x,T)/N$ is not a linear function of $\zeta$ as can be seen in Fig. \ref{fig:Ssym}. 
Namely, the deviation of $F_\mathrm{I}$ from the straight line in Fig. \ref{fig:fsym} is caused by the $x$-dependence of $S/N$.

\section{Summary}
In this paper, we have constructed an equation of state (EOS) for uniform asymmetric nuclear matter (ANM) at zero and finite temperatures using the variational method, which starts from the nuclear Hamiltonian with the AV18 and UIX potentials. 
At zero temperature, the obtained energy per nucleon $E/N$ is reasonable as compared with that obtained by APR, and the mass-radius relation of NS calculated with the present EOS is consistent with observational data. 
We also confirmed the validity of the quadratic behavior of $E/N$ with respect to the proton fraction $x$ at zero temperature.    
At finite temperature, the obtained free energy and other thermodynamic quantities such as internal energy, entropy, pressure and chemical potentials for ANM, which are important physical inputs in SN simulations, are reasonable. 
We also confirmed the thermodynamic self-consistency of the present calculations, and the validity of the frozen-correlation approximation employed in this study.    
Finally, at finite temperatures, we have found a considerable deviation of the free energy per nucleon $F/N$ from the quadratic behavior with respect to $x$, and have shown that it is caused by the $x$-dependence of the entropy. 

In order to complete a new nuclear EOS table for SN simulations, the EOS of non-uniform nuclear matter is necessary.  
For this purpose, a Thomas-Fermi calculation for non-uniform matter, with use of the values of $F/N$ for uniform matter constructed in this paper, is in progress.


\appendix
\section{Explicit expressions for the functions employed in the energy formulas}
In this appendix, we present some of the explicit expressions used in this paper.  
At zero temperature, 
$F_{\mathrm{F}ts}^{\mu}(r)$, $F_{\mathrm{qF}ts}^{\mu}(r)$ and $F_{\mathrm{bF}ts}^{\mu}(r)$ in Eqs. (\ref{eq:E2})-(\ref{eq:Fqsot}) are given as
\begin{eqnarray}
F_{\mathrm{F}ts}^{\mu}(r_{12}) &\equiv & \mathnormal{\Omega}^2{\textstyle\sum\limits_{\mathrm{isospin}}}{\textstyle\sum\limits_{\mathrm{spin}}} \int \mathnormal{\Phi}_{\mathrm{F}}^{\dagger} P_{ts12}^{\mu}\mathnormal{\Phi}_{\mathrm{F}}d\mbox{\boldmath$r$}_3d\mbox{\boldmath$r$}_4\cdots d\mbox{\boldmath$r$}_N  \nonumber  \\
    &=& \frac{2s+1}{4}\Bigg\{\xi_i\xi_j-(-1)^{t+s}l_i(r_{12})l_j(r_{12})\Bigg\},       \label{eq:FF}
\end{eqnarray}
\begin{eqnarray}
F_{\mathrm{qF}ts}^{\mu}(r_{12}) &\equiv & \mathnormal{\Omega}^2{\textstyle\sum\limits_{\mathrm{isospin}}}{\textstyle\sum\limits_{\mathrm{spin}}} \int \mathnormal{\Phi}_{\mathrm{F}}^{\dagger}\left| \mbox{\boldmath$L$}_{12} \right| ^2 P_{ts12}^{\mu}\mathnormal{\Phi}_{\mathrm{F}}d\mbox{\boldmath$r$}_3d\mbox{\boldmath$r$}_4\cdots d\mbox{\boldmath$r$}_N  \nonumber  \\
    &=& \frac{2s+1}{4}\Bigg\{\frac{r_{12}^2}{10}\xi_i\xi_j(k^2_{\mathrm{F}i}+k^2_{\mathrm{F}j})   \nonumber \\
&&-(-1)^{t+s}\frac{r_{12}}{2}\bigg[l_i(r_{12})\frac{dl_j(r_{12})}{dr_{12}}+l_j(r_{12})\frac{dl_i(r_{12})}{dr_{12}}\bigg]\Bigg\},      \label{eq:FqF}
\end{eqnarray}
\begin{eqnarray}
F_{\mathrm{bF}ts}^{\mu}(r_{12}) &\equiv & \mathnormal{\Omega}^2{\textstyle\sum\limits_{\mathrm{isospin}}}{\textstyle\sum\limits_{\mathrm{spin}}} \int \mathnormal{\Phi}_{\mathrm{F}}^{\dagger}\left| \mbox{\boldmath$L$}_{12} \right| ^4 P_{ts12}^{\mu}\mathnormal{\Phi}_{\mathrm{F}}d\mbox{\boldmath$r$}_3d\mbox{\boldmath$r$}_4\cdots d\mbox{\boldmath$r$}_N  \nonumber  \\
    &=& \frac{2s+1}{4}\Bigg\{\frac{r_{12}^4}{70}\xi_i\xi_j(k^4_{\mathrm{F}i}+k^4_{\mathrm{F}j})+\frac{r_{12}^4}{25}\xi_i\xi_jk^2_{\mathrm{F}i}k^2_{\mathrm{F}j}   \nonumber \\
&&+\frac{r_{12}^2}{5}\xi_i\xi_j(k^2_{\mathrm{F}i}+k^2_{\mathrm{F}j})-(-1)^{t+s}r_{12}^2\bigg\{\frac{dl_i(r_{12})}{dr_{12}}\frac{dl_j(r_{12})}{dr_{12}}  \nonumber  \\
&&+\frac{1}{2}\bigg[l_i(r_{12})\frac{d^2l_j(r_{12})}{dr_{12}^2}+l_j(r_{12})\frac{d^2l_i(r_{12})}{dr_{12}^2}  \nonumber \\
&&+\frac{l_i(r_{12})}{r_{12}}\frac{dl_j(r_{12})}{dr_{12}}+\frac{l_j(r_{12})}{r_{12}}\frac{d^2l_i(r_{12})}{dr_{12}^2}\bigg]\bigg\}\Bigg\}.     \label{eq:FbF}
\end{eqnarray}
Here, 
$r_{12} \equiv \left| \mbox{\boldmath$r$}_1- \mbox{\boldmath$r$}_2 \right|$, $\mathnormal{\Omega}$ is the volume of the system and $\textstyle\sum$ represents the summation over the isospin and spin coordinates of all nucleons.  
The subscripts $(i, j)$ in Eqs. (\ref{eq:FF})-(\ref{eq:FbF}) represent (p,p), (p,n) and (n,n) for $\mu=(+, 0, -)$, respectively.  
Furthermore, $(\xi_{\mathrm{p}}, \xi_{\mathrm{n}})=(x,1-x)$ and 
\begin{equation}
l_i(r)\equiv 3\xi_i\frac{j_1(k_{\mathrm{F}i}r)}{k_{\mathrm{F}i}r} \ \ \ \ \ \ \ (i=\mathrm{p}, \mathrm{n}).  
\end{equation}

At finite temperatures, the expression of $E_{\mathrm{2T}}/N$ is similar to that of $E_2/N$ given by Eq. (\ref{eq:E2}), except that $E_1/N$, $F_{\mathrm{F}ts}^{\mu}(r)$, $F_{\mathrm{qF}ts}^{\mu}(r)$ and $F_{\mathrm{bF}ts}^{\mu}(r)$ defined above are replaced by $E_{\mathrm{1T}}/N$, $F_{\mathrm{F}ts}^{\mu}(r; T)$, $F_{\mathrm{qF}ts}^{\mu}(r; T)$ and $F_{\mathrm{bF}ts}^{\mu}(r; T)$ as follows: 

\begin{equation}
\frac{E_{1\mathrm{T}}}{N}\equiv \frac{\hbar^2}{2m}\frac{1}{\pi^2\rho}\bigg[\int_0^{\infty}n_{\mathrm{p}}(k)k^4dk+\int_0^{\infty}n_{\mathrm{n}}(k)k^4dk\bigg],
\end{equation}
\begin{equation}
F_{\mathrm{F}ts}^{\mu}(r; T) \equiv \frac{2s+1}{4}\Bigg\{\xi_i\xi_j-(-1)^{t+s}l_i(r; T)l_j(r; T)\Bigg\},
\end{equation}
\begin{eqnarray}
F_{\mathrm{qF}ts}^{\mu}(r; T) &\equiv & \frac{2s+1}{4}\Bigg\{\frac{r^2}{6\pi^2\rho}\bigg[\xi_i\int_0^{\infty}n_j(k)k^4dk+\xi_j\int_0^{\infty}n_i(k)k^4dk\bigg]   \nonumber \\
&&-(-1)^{t+s}\frac{r}{2}\bigg[l_i(r;T)\frac{dl_j(r;T)}{dr}+l_j(r;T)\frac{dl_i(r;T)}{dr}\bigg]\Bigg\},
\end{eqnarray}
\begin{eqnarray}
F_{\mathrm{bF}ts}^{\mu}(r; T) &\equiv & \frac{2s+1}{4}\Bigg\{\frac{r^2}{3\pi^2\rho}
\bigg[\xi_i\int_0^{\infty}n_j(k)\bigg(k^4+\frac{r^2}{10}k^6\bigg)dk \nonumber \\
&&+\xi_j\int_0^{\infty}n_i(k)\bigg(k^4+\frac{r^2}{10}k^6\bigg)dk\bigg]  \nonumber \\
&&+\bigg[\frac{r^2}{3\pi^2\rho} \int_0^{\infty}n_i(k)k^4dk\bigg]
\bigg[\frac{r^2}{3\pi^2\rho}\int_0^{\infty}n_j(k)k^4dk\bigg]  \nonumber  \\
&&-(-1)^{t+s}r^2\bigg\{\frac{dl_i(r;T)}{dr}\frac{dl_j(r;T)}{dr}  \nonumber  \\
&&+\frac{1}{2}\bigg[l_i(r;T)\frac{d^2l_j(r;T)}{dr^2}+l_j(r;T)\frac{d^2l_i(r;T)}{dr^2}  \nonumber \\
&&+\frac{l_i(r;T)}{r}\frac{dl_j(r;T)}{dr}+\frac{l_j(r;T)}{r}\frac{dl_i(r;T)}{dr}\bigg]\bigg\}\Bigg\}.
\end{eqnarray}
Here, $l_i(r;T)$ is defined as 
\begin{equation}
l_i(r;T)\equiv \frac{1}{\pi^2\rho}\int_0^{\infty} n_i(k)j_0(kr)k^2dk \ \ \ \ \ \ \ (i=\mathrm{p}, \mathrm{n}).
\end{equation}

\section{Validity of the interpolation by Lagaris and Pandharipande for energies of asymmetric nuclear matter}
In this appendix, we examine the validity of the interpolation proposed by Lagaris and Pandharipande (LP) \cite{LP} for the calculation of $E(\rho,x)/N$ in ANM at zero temperature; this interpolation is also adopted in paper II.  
In this method, $E(\rho,x)/N$ is approximated with use of $E(\rho,x)/N$ for SNM and PNM as follows:
\begin{equation}
\frac{E_{\mathrm{IntLP}}(\rho, x)}{N}=\frac{E_1(\rho,x)}{N}+\frac{E_{\mathrm{V}}(\rho, 1/2)}{N}+\Bigg[\frac{E_{\mathrm{V}}(\rho, 0)}{N}-\frac{E_{\mathrm{V}}(\rho, 1/2)}{N}\Bigg](1-2x)^2.   \label{eq:Type2}
\end{equation}
Here, $E_1(\rho,x)/N$ is the one-body kinetic energy given in Eq. (\ref{eq:FermiE}), and
\begin{equation}
\frac{E_{\mathrm{V}}(\rho, x)}{N}=\frac{E(\rho, x)}{N} - \frac{E_1(\rho, x)}{N}.   \label{eq:EInter}
\end{equation}
In Fig. \ref{fig:delta0}, we show the relative deviation of $E_{\mathrm{IntLP}}(\rho, x)/N$ defined by an equation similar to Eq. (\ref{eq:deltaE}), i.e., 
\begin{equation}
\mathnormal{\Delta} E_\mathrm{LP}(\rho, x) = \frac{E_{\mathrm{IntLP}}(\rho, x)/N - E(\rho, x)/N}{E_{\mathrm{I}}(\rho,x)}.   \label{eq:deltaELP}
\end{equation}
It can be seen that $|\mathnormal{\Delta} E_\mathrm{LP}(\rho, x)|$ is much smaller than that of the usual interpolation Eq. (\ref{eq:Type1}), because $E_1(\rho,x)/N$, which is the main cause of the relatively large values of $\mathnormal{\Delta} E$, is removed from the interpolated part: 
The interpolation by LP is a good approximation for ANM at zero temperature.

\section*{Acknowledgements}
We would like to express special thanks to H. Kanzawa, K. Nakazato, K. Oyamatsu, Y. Sekiguchi, K. Sumiyoshi, H. Suzuki, M. Yamada, S. Yamada and N. Yasutake for valuable discussions and comments, to A. Mukherjee for providing us with numerical data on free energies for the FHNC calculation, 
and to H. Matsufuru for supporting on the parallel computing. 
A part of the numerical computations in this work was performed on the GPGPU at the High Energy Accelerator Research Organization (KEK). 
This work is supported by JSTS (No. 21540280), JSPS Research Fellowships for Young Scientists (No. 24-3275), and the Grant-in-Aid for Scientific Research on Innovative Areas of MEXT (Nos. 20105003, 20105004, 20105005).  


\bibliographystyle{model1-num-names}
\bibliography{<your-bib-database>}



\end{document}